\documentclass[prb,aps,twocolumn,footinbib,groupedaddress,citeautoscript]{revtex4-1}

\usepackage{amssymb}
\usepackage{amsmath}
\usepackage{amsfonts}
\usepackage{color}
\usepackage{graphicx}
\usepackage{bm}

\usepackage[unicode]{hyperref}
\hypersetup{
   unicode=true,          
   plainpages=false,
   colorlinks=true,       
   citecolor=blue,        
}

\usepackage{array}
\newcolumntype {s}[1]{@{\hspace{#1}}} 
\newcolumntype {R}{>{$}r<{$}}         
\newcolumntype {C}{>{$}c<{$}}         
\newcolumntype {L}{>{$}l<{$}}         
\newcolumntype {f}{@{\extracolsep\fill}}  

\newcommand* {\vek}[1]{{\ensuremath{\bm{\mathrm{#1}}}}}
\newcommand* {\kk}{\vek{k}}
\newcommand* {\rr}{\vek{r}}

\newcommand* {\bra}[1]{\ensuremath{\langle {#1} |}}
\newcommand* {\ket}[1]{\ensuremath{| {#1} \rangle}}

\newcommand* {\tvek}[2][c]{\left( \begin{array}{s{0.15em}#1s{0.15em}}
     #2\end{array} \right)}
\newcommand* {\ee}{\ensuremath{\mathrm{e}}}
\newcommand* {\ham}{\mathsf{H}}

\graphicspath{{.}{./EPS/}}

\begin{document}

\title{Dirac electrons in quantum rings}

\author{L. Gioia}
\affiliation{School of Chemical and Physical Sciences and MacDiarmid
Institute for Advanced Materials and Nanotechnology, Victoria
University of Wellington, PO Box 600, Wellington 6140, New Zealand}

\author{U. Z\"ulicke}
\email{uli.zuelicke@vuw.ac.nz}
\affiliation{School of Chemical and Physical Sciences and MacDiarmid
Institute for Advanced Materials and Nanotechnology, Victoria
University of Wellington, PO Box 600, Wellington 6140, New Zealand}

\author{M. Governale}
\affiliation{School of Chemical and Physical Sciences and MacDiarmid
Institute for Advanced Materials and Nanotechnology, Victoria
University of Wellington, PO Box 600, Wellington 6140, New Zealand}

\author{R. Winkler}
\affiliation{Department of Physics, Northern Illinois University,
DeKalb, Illinois 60115, USA}
\affiliation{Materials Science Division, Argonne National Laboratory,
Argonne, Illinois 60439, USA}

\date{\today}

\begin{abstract}

We consider quantum rings realized in materials where the dynamics
of charge carriers mimics that of two-dimensional (2D) Dirac
electrons. A general theoretical description of the ring-subband
structure is developed that applies to a range of currently available
2D systems, including graphene, transition-metal dichalcogenides,
and narrow-gap semiconductor quantum wells. We employ the
scattering-matrix approach to calculate the electronic two-terminal
conductance through the ring and investigate how it is affected by
Dirac-electron interference. The interplay of pseudo-spin chirality
and hard-wall confinement is found to distinctly affect the geometric
phase that is experimentally accessible in mesoscopic-conductance
measurements. We derive an effective Hamiltonian for the azimuthal
motion of charge carriers in the ring that yields deeper insight into
the physical origin of the observed transport effects, including the
unique behavior exhibited by the lowest ring subband in the normal and
topological (i.e., band-inverted) regimes. Our work provides a unified
approach to characterizing confined Dirac electrons, which can be used
to explore the design of valley- and spintronic devices based on
quantum interference and the confinement-tunable geometric phase.

\end{abstract}

\maketitle

\section{Introduction}
\label{sec:intro}

Quantum rings~\cite{fom14} are a paradigmatic system for studying
topological effects in condensed matter. In particular, coherent
electron transport through ballistic rings and similar multiply
connected conductors can be used to reveal phenomena associated
with geometric phases~\cite{ber84, aha87}, including the
Aharonov-Bohm~\cite{aha59, gef84, bue84} and
Aharonov-Casher~\cite{aha84, aro93, qia94, kov07} effects as well as
non-Abelian generalizations~\cite{wil84, aro98}. Besides the coupling
of charge carriers to effective gauge fields, quantum confinement in
the ring structure turns out to also importantly affect
coherent-electron interference~\cite{bue85, fru04, ple08}, which
further increases possibilities for its experimental control and
application for novel electronic-device functionalities.

Our present work is motivated by the recent interest in
two-dimensional (2D) materials with Dirac-like charge carriers such
as single-layer graphene~\cite{nov05, zha05, cas09}, single-layer
transition-metal dichalcogenides~\cite{xia12, kor15}, and quantum
wells in narrow-gap semiconductors~\cite{whi81, vol85} such as
HgTe~\cite{ber06, kon07} and InAs/GaSb~\cite{liu08, du17}. These
condensed-matter realizations of 2D Dirac electrons necessarily carry
a two-valued flavor degree of freedom~\footnote{The valley isospin of
massless Dirac electrons in graphene constitutes an example for such a
flavor quantum number, as does the real spin of charge carriers in a
semiconductor quantum well. The two-flavor 2D-Dirac model was
originally introduced in the context of quantum field
theory~\cite{jac81}. Its application in condensed-matter physics has
recently been reviewed in Refs.~\onlinecite{gus07, win15a}.}.
Ring structures in single-layer graphene have previously been studied
by analytical and numerical solution of continuum-model-based Dirac
equations~\cite{zar10, sti13, bol14, bel16, rec07, dco14, wur10} and
also numerical tight-binding calculations~\cite{rec07, wur10, sch12,
rom13, dco14}. The bound states in a ring conductor realized in
narrow-gap semiconductor quantum wells were also
considered~\cite{gon08, mic11}. Very recently, a theoretical study of
quantum rings in MoS$_2$ has been performed~\cite{oli16}.
Experimental realizations have been achieved in HgTe/HgCdTe quantum
wells~\cite{kon06}, graphene~\cite{rus08, yoo10, hue10, smi12, dau17}
and MoS$_2$~\cite{fan14}.

In contrast to previous theoretical studies that have largely
focused on the specifics of various materials systems, we present a
broadly applicable and systematic description of the electronic
structure and quantum-interference effects in 2D-Dirac-electron
quantum-ring conductors based on a completely general
subband-$\kk\cdot\vek{p}$ approach. We obtain an effective
Hamiltonian for the azimuthal motion of ring-confined Dirac-like
charge carriers that provides deeper insight into characteristic
features of the electronic subband structure and allows to explore
physical implications for quantum-transport effects. Complementing
existing work that has largely focused on persistent currents in
isolated Dirac rings~\cite{rec07, zar10, sti13, bol14, bel16, mic11}
or studied transport through a particular Dirac-ring realization
numerically~\cite{wur10, sch12}, we present analytic results for the
two-terminal conductance. In typical experiments~\cite{kon06, rus08, 
yoo10, hue10, smi12, dau17} and previous numerical
studies~\cite{wur10, sch12}, the entire structure consisting of the
ring conductor and external leads was made out of the same
material. This motivated us to discuss in detail the case of
flavor-conserving scattering of Dirac electrons at the ring-lead
junctions. We identify a purely confinement-induced contribution to
the geometric phase, which turns out to have opposite sign for the
two flavors of 2D Dirac electrons propagating in the ring. We use
this observation to explore possible uses of ring conductors as
\emph{flavortronic\/} devices.\footnote{Depending on the physical
origin of the flavor degree of freedom in particular materials,
\emph{flavortronics} can be synonymous with
\emph{valleytronics}~\cite{sch16} (e.g., in graphene) or
spintronics~\cite{sin12} (e.g., in semiconductor quantum wells) or
a combination of both (e.g., in transition-metal
dichalchogenides~\cite{xu14}).}

The remainder of this Article is organized as follows. We start by
introducing the generic model Hamiltonian describing two-flavor
2D-Dirac electrons in a variety of materials in Sec.~\ref{sec:2Dham}.
The general subband-$\kk\cdot\vek{p}$ description of Dirac-electron
rings is developed in Sec.~\ref{sec:model}. As part of the derivation,
the radial hard-wall-confinement problem for Dirac electrons is solved
(Sec.~\ref{sec:hardWall0}) and an effective Hamiltonian for the
azimuthal motion of charge carriers in the ring is obtained
(Sec.~\ref{sec:effAzimHam}). Armed with the understanding of
Dirac-ring subband structure, Sec.~\ref{sec:transport} discusses how
quantum-interference effects are exhibited in the conductance through
the ring. Our scattering-matrix approach is introduced in
Sec.~\ref{sec:Tjunc}, and the fully general two-terminal transmission
function for a clean ring with flavor-conserving scattering at the
ring-lead junctions is presented. Possible applications of Dirac rings
as flavortronics devices are explored in Sec.~\ref{sec:condRes}. We
summarize our conclusions in Sec.~\ref{sec:concl}, and relevant
mathematical details are given in the Appendices.

\section{Two-flavor 2D-Dirac Hamiltonian}
\label{sec:2Dham}

The motion of electrons in 2D materials is described by an
envelope-function Hamiltonian that can be written in the generic
two-flavor 2D-Dirac form~\footnote{The form of the $4\times 4$
Hamiltonian (\ref{eq:4x4Ham}) presumes an appropriate choice of basis
functions, and the relation between its diagonal $2\times 2$ blocks
(the vanishing of its off-diagonal $2\times 2$ blocks) is due to
time-reversal (parity) symmetry. See, e.g., Ref.~\onlinecite{faj17} for a
more detailed overview of specific physical realizations.}
\begin{subequations}
\begin{eqnarray}\label{eq:4x4Ham}
H &=& \left( \begin{array}{cc} \mathcal{H}(\kk) & 0 \\ 0 &
\mathcal{H}^\ast(-\kk) \end{array} \right) \equiv \left(
\begin{array}{cc} \mathcal{H}^{(+)} & 0 \\ 0 & \mathcal{H}^{(-)}
\end{array} \right)\,\, , \\ \label{eq:univHam}
\mathcal{H}^{(\pm)} &=& \pm\gamma \left( k_\pm \, \sigma_- + k_\mp\,
\sigma_+ \right) + \frac{\Delta(k)}{2} \, \sigma_3 + \epsilon(k)\, \sigma_0
\, , \quad
\end{eqnarray}
\end{subequations}
where $\sigma_\pm = (\sigma_1 \pm i\, \sigma_2)/2$ are ladder
operators for the eigenstates of the diagonal Pauli matrix $\sigma_3$
that correspond to the $\kk = 0$ conduction and valence-band states,
$\sigma_0$ is the $2\times 2$ identity matrix, and $k_\pm := k_x \pm
i\, k_y$ in terms of Cartesian components of the in-plane wave vector
$\kk\equiv (k_x, k_y)$. The parameter $\gamma$ characterizes the
inter-band coupling, and the gap and electron-hole-asymmetry terms
are of the general form
\begin{subequations}
\begin{eqnarray}\label{eq:delta}
\Delta(k) &=& \Delta_0 + \frac{2 \gamma}{k_\Delta} \, k^2 \quad ,
\\[0.1cm]\label{eq:asymm}
\epsilon(k) &=& \epsilon_0 + \xi\, \frac{\gamma}{k_\Delta}
\, k^2 \quad ,
\end{eqnarray}
\end{subequations}
with contributions quadratic in $k$ arising generically due to the
influence of remote bands~\cite{whi81, win03}. The parameter
$k_\Delta$ is the wave-vector scale at which remote-band
contributions $\propto k^2$ to the gap become comparable to the
inter-band coupling $\propto \gamma k$, and the dimensionless
number $\xi$ is a measure of broken electron-hole symmetry. As
$\epsilon_0$ constitutes an irrelevant uniform shift in energy, we set
$\epsilon_0=0$ for convenience. The values of parameters in the
Hamiltonian (\ref{eq:univHam}) for specific materials are given in
Table~\ref{table:parameter}. Systems with $\Delta_0 > 0$ are ordinary,
i.e., nontopological, insulators. In contrast, $\Delta_0 < 0$
signifies the band inversion occurring in topological
insulators~\cite{fra13}.

\begin{table}[b]
\caption{\label{table:parameter}
Parameters in the effective 2D-Dirac Hamiltonians for electrons
in some representative single-layer (SL) atomic crystals and
semiconductor quantum wells (QW).} 
\renewcommand{\arraystretch}{1.1}
\begin{tabular*}{\columnwidth}{lfcccc}
\hline \hline \rule{0pt}{2.5ex}
& $\gamma$ (eV\AA) & $\Delta_0$ (eV) & $k_\Delta$ (\AA$^{-1}$) & 
$\xi$ \\ \hline
SL graphene\footnote{Refs.~\onlinecite{cas09, gru08}} & 6.4 &
$\lesssim 0.01$ & 0.17 & 0.026 \\
SL MoS$_2$\footnote{Ref.~\onlinecite{kor15}} & 3.0 & 1.7 & 0.91
& 0.89 \\
HgTe/CdTe QW\footnote{Ref.~\onlinecite{fra13}, p.~64 (HgTe well
width: $7.0\,$nm)} & 3.7 & $-0.020$ & 0.053 & 0.74 \\
InAs/GaSb QW\footnote{Ref.~\onlinecite{fra13}, p.~65 (InAs/GaSb
well widths: $10\,$nm/$10\,$nm)}  & 0.37 & $-0.016$ & 0.0056 &
0.088 \\ \hline \hline
\end{tabular*}
\end{table}

Switching to polar coordinates $\rr = (r, \varphi)$, we take $\kk
\equiv i \vek{\nabla}$ to be an operator in real-space representation
and note the relation
\begin{equation}
k_\pm = \ee^{\pm i \varphi / 2} \left( k_r \pm i\, k_\varphi \right)
\, \ee^{\pm i \varphi / 2} \quad ,
\end{equation}
with the Hermitian operators~\cite{paz01}
\begin{subequations}
\begin{eqnarray}
k_r &=& -i \left( \partial_r + \frac{1}{2 r} \right)\quad , \\[0.1cm]
k_\varphi &=& - i\, \frac{\partial_\varphi}{r}  \quad .
\end{eqnarray}
\end{subequations}
As the Hamiltonians $\mathcal{H}^{(\pm)}$ commute with total angular
momentum $J_z^{(\pm)} = -i\hbar \, \sigma_0 \, \partial_\varphi \pm \hbar
\, \sigma_3/2$, it is useful to switch to a representation of diagonal
$J_z^{(\pm)}$ using the transformation
\begin{equation}
 \mathcal{U}_\pm(\varphi) = \exp (\mp i \,\sigma_3 \, \varphi/2 ) \quad .
\end{equation}
It is straightforward to obtain 
\begin{equation}
\mathcal{H}^{(\tau)} = \mathcal{U}_\tau(\varphi) \left( \mathcal{H}^{(\tau)}_r +
\mathcal{H}^{(\tau)}_\varphi \right) \mathcal{U}_\tau^\dagger(\varphi)
\quad ,
\end{equation}
where $\tau=\pm$ labels the two flavors of 2D-Dirac electrons, and 
\begin{subequations}
\label{eq:Ham:polar}
\begin{eqnarray}
\mathcal{H}^{(\tau)}_r &=& \tau\, \gamma \, k_r\, \sigma_1 +
\frac{\Delta_0}{2} \, \sigma_3 + \frac{\gamma}{k_\Delta} \left(
\sigma_3 + \xi\, \sigma_0 \right) k_r^2 \,\, , \label{eq:Ham:polar:rad}\\
\mathcal{H}^{(\tau)}_\varphi &=& \gamma \, k_\varphi \, \sigma_2 +
\frac{\gamma}{k_\Delta} \left( \sigma_3 + \xi\, \sigma_0 \right)\left(
k_\varphi^2 - \tau \, \sigma_3 \, \frac{k_\varphi}{r} \right) \quad
\label{eq:Ham:polar:az}\end{eqnarray}
\end{subequations}
describe their motion in radial and azimuthal coordinates. The
expressions (\ref{eq:Ham:polar}) form the basis for our further
study of quantum states in ring conductors.

\section{Ring-confined Dirac electrons}
\label{sec:model}

We assume the ring structure to be defined by an axially symmetric
mass confinement~\footnote{We follow previous works on ring
structures in 2D-Dirac materials~\cite{rec07, mic11} where a mass
confinement embodied by the Lorentz-scalar potential~\cite{cou88}
$H_V$ given in Eq.~(\ref{eq:massConf}) was adopted to simulate
confinement realized by lithographic techniques~\cite{rus08, yoo10,
hue10, smi12, dau17, fan14}.  As a possible alternative, a radially
symmetric Lorentz-\emph{vector} potential~\cite{cou88}
$$ H_U = \left( \begin{array}{cc} U(r)\,\sigma_0 & 0 \\[0.3cm] 0 &
    U(r)\, \sigma_0\end{array} \right) $$
could be considered. Such a potential, which generally arises from
electrostatic confinement, was used previously to model gate-defined
graphene quantum dots. See, e.g., Refs.~\onlinecite{sil07, mat08, rec09}.}
\begin{equation}\label{eq:massConf}
H_V = \left( \begin{array}{cc} V(r)\,\sigma_3 & 0 \\[0.3cm] 0 &
V(r)\,\sigma_3 \end{array} \right) \quad .
\end{equation}
Because of the axial symmetry of the potential $V(r)$, the $2\times 2$
Schr\"odinger equations
\begin{subequations}
\begin{equation}\label{eq:Schroed1}
\left[ \mathcal{H}^{(\tau)} + V(r)\,\sigma_3 \right]
\ket{\Psi^{(\tau)}} = E \, \ket{\Psi^{(\tau)}}
\end{equation}
can be written as
\begin{equation}\label{eq:Schroed2}
\mathcal{U}_\tau(\varphi) \left[ \mathcal{H}^{(\tau)}_r +
\mathcal{H}^{(\tau)}_\varphi + V(r)\,\sigma_3 \right]
\mathcal{U}_\tau^\dagger(\varphi)\, \ket{\Psi^{(\tau)}} = E \,
\ket{\Psi^{(\tau)}} \,\, ,
\end{equation}
\end{subequations}
motivating the separation \textit{Ansatz\/}
\begin{equation}\label{eq:Ansatz}
\ket{\Psi^{(\tau)}} = \ee^{i l \varphi} \,\, \mathcal{U}_\tau
(\varphi) \, \frac{1}{\sqrt{2\pi r}} \, \ket{\Phi_l^{(\tau)}} \quad ,
\end{equation}
where we introduced the azimuthal quantum number $l$.
This transforms Eq.~(\ref{eq:Schroed2}) into the Schr\"odinger
equation for a confined Dirac particle in one spatial
dimension~\cite{whi81, ber87, mck87, cou88, alb96},
\begin{equation}\label{eq:confDirac}
\left[ \mathcal{H}_\mathrm{1D}^{(\tau)} + V(r)\,\sigma_3 +
\mathcal{V}^{(\tau)}_l(r) \right] \ket{\Phi_l^{(\tau)}} = E_l^{(\tau)} \,
\ket{\Phi_l^{(\tau)}} \quad ,
\end{equation}
with
\begin{subequations}
\begin{equation}
\mathcal{H}_\mathrm{1D}^{(\tau)} =  -i \tau\, \gamma\, \sigma_1
\, \frac{d}{dr} + \frac{\Delta_0}{2}\,\sigma_3 -
\frac{\gamma}{k_\Delta}\left( \sigma_3 + \xi\, \sigma_0
\right) \frac{d^2}{dr^2}
\end{equation}
and the centrifugal-barrier contribution
\begin{equation}
\mathcal{V}^{(\tau)}_l(r) = \gamma \, \frac{l}{r}\left\{ \sigma_2 +
\frac{1}{k_\Delta r} \left[ l \left( \sigma_3 + \xi\, \sigma_0 \right) -
\tau \left( \sigma_0 + \xi\, \sigma_3 \right) \right] \right\} .
\end{equation}
\end{subequations}

In the spirit of subband-$\vek{k}\cdot\vek{p}$ theory~\cite{bro85a, 
bro85b}, we start by considering Eq.~(\ref{eq:confDirac}) for
$l=0$,\footnote{States with arbitrary value of $l$, specifically also
$l=0$, are physical when the quantum ring is connected to leads,
which is the situation we are interested in here. In contrast, for an
isolated ring, $l$ is required to be \emph{half-integer\/} for physical
states.}
\begin{equation}\label{eq:l=0SE}
\left[ \mathcal{H}_\mathrm{1D}^{(\tau)} + V(r)\,\sigma_3 \right]
\ket{\Phi_0^{(\tau,n)}} = E_0^{(\tau,n)} \, \ket{\Phi_0^{(\tau,n)}}
\quad ,
\end{equation}
and then use the eigenstates $\ket{\Phi_0^{(\tau,n)}}$ as a new basis
to calculate the ring-subband dispersions $E_l^{(\tau,n)}$. Here the
radial quantum number $n= \pm 1, \pm 2, \dots$ labels the ring
subbands with the usual convention $E_0^{(\tau,n)} > E_0^{(\tau,n')}$
for $n>n'$. A general eigenstate with $l\ne 0$ is thus expressed as a
superposition of basis states,
\begin{equation}\label{eq:kdotpBasis}
\ket{\Phi_l^{(\tau, n)}} = \sum_{n' > 0} \left( a_{l n'}^{(\tau, n)}
\, \ket{\Phi_0^{(\tau, n')}} + b_{l n'}^{(\tau, n)} \,
\ket{\Phi_0^{(\tau,-n')}} \right) ,
\end{equation}
with coefficients $a_{l n'}^{(\tau, n)}$ and $b_{l n'}^{(\tau, n)}$
that need to be determined by solving the eigenvalue equation
\begin{equation}\label{eq:eigenvalueeq}
\ham_l^{(\tau)} \tvek{a_{l 1}^{(\tau, n)} \\ b_{l 1}^{(\tau, n)} \\
a_{l 2}^{(\tau, n)} \\ b_{l 2}^{(\tau, n)} \\  \vdots} = E_l^{(\tau,
n)} \, \tvek{a_{l 1}^{(\tau, n)} \\ b_{l 1}^{(\tau, n)} \\ a_{l
2}^{(\tau, n)} \\ b_{l 2}^{(\tau, n)} \\  \vdots} \quad ,
\end{equation}
with the new Hamiltonian matrix
\begin{subequations}
\begin{equation}\label{eq:xLham}
\ham_l^{(\tau)} =
\left( \begin{array}{ccc}
\big( \ham_l^{(\tau)} \big)_{1, 1} & \big( \ham_l^{(\tau)}
\big)_{1, 2} & \ldots \\[0.3cm] \big( \ham_l^{(\tau)} \big)_{2, 1} &
\big( \ham_l^{(\tau)} \big)_{2, 2} & \ldots \\[0.3cm] \vdots & \vdots
& \ddots \end{array} \right)
\end{equation}
whose $2\times 2$ sub-blocks are given by
\begin{widetext}
\begin{equation}\label{eq:2x2Lham}
\big( \ham_l^{(\tau)} \big)_{n, n'} = \left( \begin{array}{cc}
E_0^{(\tau,n)} \delta_{n n'} + \big\langle \mathcal{V}^{(\tau)}_l(r) \big 
\rangle^{(\tau)}_{n,n'} & \big\langle \mathcal{V}^{(\tau)}_l(r) \big
\rangle^{(\tau)}_{n,-n'} \\[0.3cm] \big\langle \mathcal{V}^{(\tau)}_l(r)
\big\rangle^{(\tau)}_{-n,n'} & E_0^{(\tau,-n)}\delta_{n n'} +
\big\langle \mathcal{V}^{(\tau)}_l(r) \big\rangle^{(\tau)}_{-n,-n'} 
\end{array}\right)\quad .
\end{equation}
\end{widetext}
\end{subequations}
Here $\langle\mathcal{O}\rangle^{(\tau)}_{n, n'} \equiv
\bra{\Phi_0^{(\tau, n)}}\mathcal{O}\ket{\Phi_0^{(\tau, n')}}$ for
any operator $\mathcal{O}$, and $\delta_{n n'}$ is the Kronecker
symbol.

In the electron-hole-symmetric case (i.e., when $\xi=0$),
the energy-reflection symmetry~\cite{win15a}
\begin{equation}
\sigma_2\left[ \mathcal{H}_{\mathrm{1D},\xi=0}^{(\tau)} +
V(r) \,\sigma_3 \right]\sigma_2 = - \left[ \mathcal{H}_{\mathrm{1D},
\xi=0}^{(\tau)} + V(r)\, \sigma_3 \right]
\end{equation}
holds, implying the relations
\begin{subequations}
\begin{eqnarray}
E_{0,\xi=0}^{(\tau,-n)} &=& - E_{0,\xi=0}^{(\tau,n)}
\quad , \\[0.1cm]
\ket{\Phi_{0,\xi=0}^{(\tau, -n)}} &=& \sigma_2\,
\ket{\Phi_{0,\xi=0}^{(\tau, n)}} \quad .
\end{eqnarray}
\end{subequations}
As a result, all matrix elements in Eq.~(\ref{eq:2x2Lham}) can then be
expressed in terms of matrix elements between eigenstates for
positive energies with labels $n, n' > 0$,
\begin{subequations}
\begin{eqnarray}
\langle\mathcal{O}\rangle^{(\tau)}_{n, -n'} &\stackrel{\xi=
0}{\longrightarrow}& \langle\mathcal{O}\sigma_2\rangle^{(\tau)}_{n,
n'} \quad , \\
\langle\mathcal{O}\rangle^{(\tau)}_{-n, n'} &\stackrel{\xi=
0}{\longrightarrow}& \langle\sigma_2\mathcal{O}\rangle^{(\tau)}_{n,
n'} \quad , \\
\langle\mathcal{O}\rangle^{(\tau)}_{-n, -n'} &\stackrel{\xi
=0}{\longrightarrow}& \langle\sigma_2\mathcal{O}\sigma_2
\rangle^{(\tau)}_{n, n'} \quad ,
\end{eqnarray}
\end{subequations}
which simplifies further calculations.

In general, Eq.~(\ref{eq:eigenvalueeq}) can be solved only
numerically. It turns out, however, that a hierarchy of relative
importance emerges among the $2\times 2$ sub-blocks in
Eq.~(\ref{eq:xLham}) in the limit of narrow rings, which can be
exploited to obtain useful approximate analytical results. We develop
this approach in the following using the specific situation of a
hard-wall confinement. As a first step, the $l=0$ eigenstates are
determined, as discussed in Sec.~\ref{sec:hardWall0}. We then
use these states as basis states for
calculating the $l\ne 0$ eigenstates of a hard-wall-confined ring
structure according to the procedure outlined formally in
Eqs.~(\ref{eq:kdotpBasis}) and (\ref{eq:eigenvalueeq}). Identification
of the most important couplings in Eq.~(\ref{eq:xLham}) then yields an
effective model for the azimuthal motion of ring-confined two-flavor
2D-Dirac electrons given in Sec.~\ref{sec:effAzimHam}.

\subsection{Hard-wall-confined quantum ring: \texorpdfstring{$\bm{l=0}
$}{} states}\label{sec:hardWall0}

To be specific, we now assume a hard-wall potential
\begin{equation}\label{eq:hardwall}
V(r) = \left\{ \begin{array}{cl} 0 & \mbox{for } R - \frac{W}{2} < r
< R + \frac{W}{2} \\[0.2cm] \infty & \mbox{elsewhere} \end{array}
\right. \quad ,
\end{equation}
where $W$ and $R$ denote, respectively, the ring's width and average
radius. We find the $l=0$ eigenstates, i.e., solutions of
Eq.~(\ref{eq:l=0SE}), for this potential by forming a general
superposition of same-energy eigenstates of
$\mathcal{H}_\mathrm{1D}^{(\tau)}$ and applying hard-wall boundary
conditions at the inner and outer ring radii. See
Appendix~\ref{sec:appA} for details of the calculation.

\begin{figure}[b]
\includegraphics[width=0.9\columnwidth]{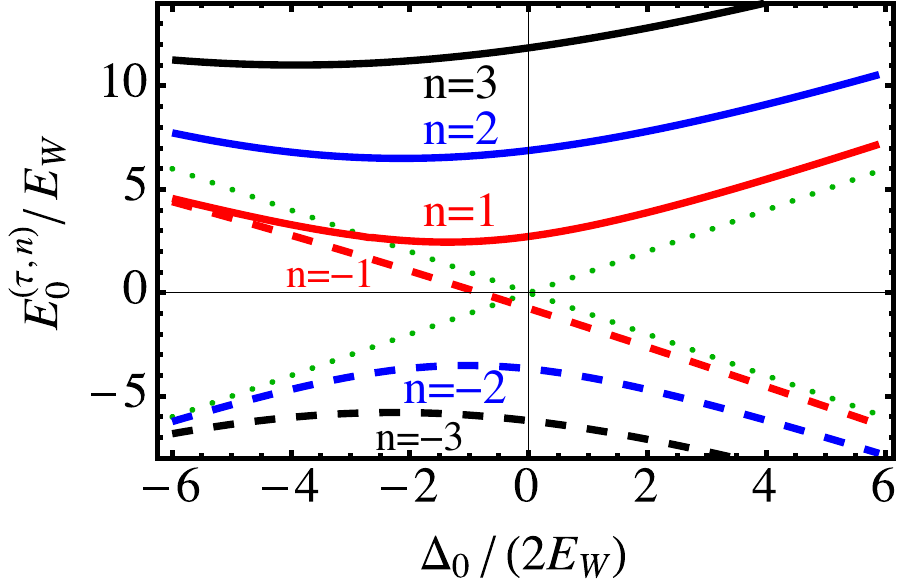}
\caption{\label{fig:BHZsubbs}%
Spectrum of $l=0$ bound-state energies for a hard-wall-confined ring
structure as a function of the 2D-Dirac gap parameter $\Delta_0$
measured in units of the size-quantization energy $E_W=\gamma/W$.
Solid (dashed) red, blue and black curves correspond to subbands with
$n = 1$, $2$, and $3$ ($n=-1$, $-2$, and $-3$), respectively. The thin
dotted green lines indicate the position of the 2D-Dirac gap edges
$\pm |\Delta_0|/2$. Except for $\Delta_0$, band-structure parameters
used in the calculation were fixed at values applicable to a 7-nm
HgTe quantum well~\cite{rot10}, and we set $k_\Delta W = 26.6$. [For
reference, the 7-nm HgTe quantum-well gap satisfies
$\Delta_0/(2 E_W) = -5.48$.] All energy levels are two-fold
degenerate in the flavor degree of freedom distinguished by
$\tau=\pm$. The $n=\pm 1$ levels lying below the 2D-Dirac gap edges
constitute hybridized quantum-spin-Hall edge states.}
\end{figure}

\begin{figure}[b]
\includegraphics[width=0.85\columnwidth]{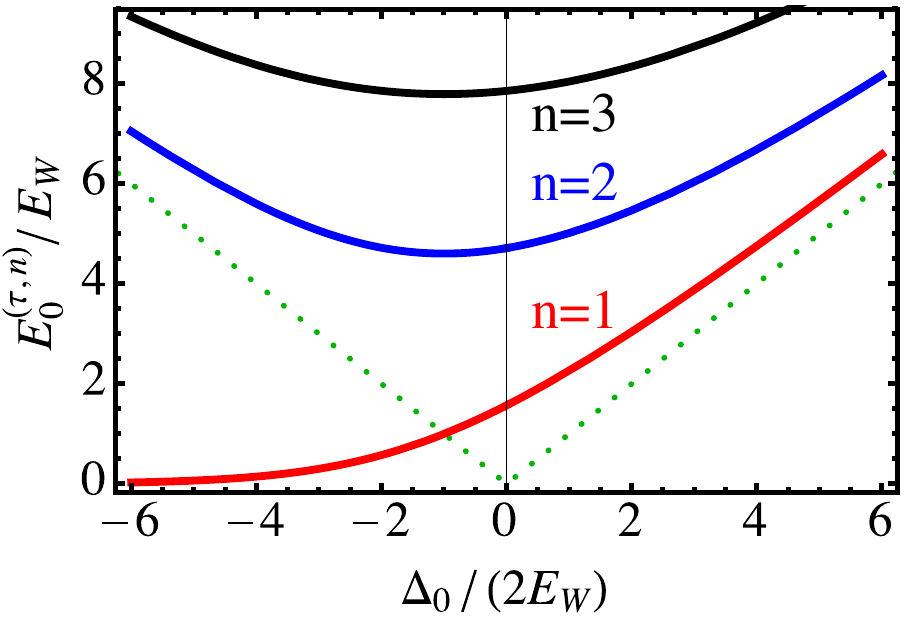}
\caption{\label{fig:SLGuniv}%
Spectrum of $l=0$ bound-state energies for a hard-wall-confined ring
structure in the ordinary-Dirac limit [$\Delta(k)=\Delta_0$ and
$\epsilon(k)=0$] as a function of the 2D-Dirac gap parameter
$\Delta_0$ measured in units of the size-quantization energy $E_W=
\gamma/W$. Red, blue and black curves correspond to subbands with
$n = 1$, $2$, and $3$, respectively.  The dotted lines indicate $E =
|\Delta_0|/2$, revealing the evanescent character of the lowest-energy
state for $\Delta_0/2 < -E_W$. All levels are two-fold degenerate in
the flavor quantum number $\tau$.}
\end{figure}

Figure~\ref{fig:BHZsubbs} illustrates the dependence of the $l=0$ ring
energies on the gap parameter $\Delta_0$, with the latter normalized
to the size-quantization energy $E_W=\gamma/W > 0$. Identical energies
are obtained for the two series of bound states distinguished by the
flavor quantum number $\tau=\pm$. The $n=\pm 1$ subbands behave
qualitatively differently from the other subbands (those having
$|n|>1$) in that they can lie below the 2D-Dirac gap edges for
sufficiently negative values of $\Delta_0$, in which case they
correspond to hybridized quantum-spin-Hall edge states~\cite{zho08}.
In particular, the energy gap between the lowest conduction ($n=+1$)
and valence ($n=-1$) subbands vanishes in the limit $-\Delta_0 \gg
E_W$. As we will see below, the low-energy electron dynamics in this
limit turns out to be ultrarelativistic, massless-1D-Dirac-like. In
contrast, subbands with $|n|>1$ have gapped energy dispersions that
satisfy $E_0^{(\tau, \pm n)}\gtrless \pm \max \{ E_W, |\Delta_0|/2\}$
and therefore exhibit nonrelativistic, ordinary-Schr\"odinger-like
behavior in sufficiently narrow rings for any value of $\Delta_0$. 

The situation simplifies considerably in the ordinary-2D-Dirac limit
where $\Delta(k)\to\Delta_0$ and $\epsilon(k)\to 0$. Firstly, the
subband energies for $l=0$ become electron-hole-symmetric; i.e.,
$E_0^{(\tau, n)} = - E_0^{(\tau, -n)}$. Secondly, the bound-state
energies measured in units of $E_W$ have a universal dependence
on $\Delta_0/(2 E_W)$. Figure~\ref{fig:SLGuniv} shows pertinent
results for the $n=1$, $2$ and $3$ subband states. All energy levels
are again two-fold degenerate in the flavor quantum number $\tau$.
The lowest-subband state becomes evanescent~\cite{cou88} for
$\Delta_0/2 < -E_W$, indicating that the system is
topological~\cite{fra13}. The fact that the transition to the inverted
regime in the quantum-ring system occurs only for sufficiently
negative values of the 2D-Dirac gap $\Delta_0$ provides another,
particularly clean, example for how size quantization generally
competes with the band inversion in topologically nontrivial
systems~\cite{liu10, kot17}.

\subsection{Effective Hamiltonian for the azimuthal motion}
\label{sec:effAzimHam}

Having derived the basis states applicable to a hard-wall quantum-ring
confinement, the form of the Hamiltonian matrix (\ref{eq:xLham}) can
be analyzed in greater detail. The diagonal $2\times 2$ sub-blocks
$\big( \ham_l^{(\tau)} \big)_{n, n}$ essentially represent azimuthal
dynamics involving only the two subbands labeled by $\pm n$ for fixed
$n>0$. In contrast, the off-diagonal $2\times 2$ sub-blocks $\big(
\ham_l^{(\tau)} \big)_{n, n'}$ with $n\ne n'$ embody the coupling
between such pairs of subbands. Here we discuss the particular form of
both kinds of sub-blocks for hard-wall-confined quantum rings.

A detailed consideration (see Appendix~\ref{sec:appB}) motivates the
parameterization of diagonal sub-blocks in Eq.~(\ref{eq:xLham}) for
a hard-wall ring confinement in the general form of a Hamiltonian
governing azimuthal motion. It can be written as the sum of two parts,
\begin{equation}\label{eq:aziHam}
\big( \ham_l^{(\tau)} \big)_{n, n} = \big( \mathsf{K}_l^{(\tau)}
\big)_{n,n} + \big( \mathsf{L}_l^{(\tau)} \big)_{n, n} \quad ,
\end{equation}
such that the part $\big( \mathsf{K}_l^{(\tau)} \big)_{n, n}$ contains
the most relevant leading terms while corrections, e.g., due to
electron-hole asymmetry, are subsumed into $\big(
\mathsf{L}_l^{(\tau)} \big)_{n,n}$. Making the leading dependences on
the ring aspect ratio $W/R$ as well as on 2D-Dirac flavor $\tau$ and
electron-hole asymmetry $\xi$ explicit, we write
\begin{widetext}
\begin{subequations}
\begin{eqnarray}\label{eq:aziHam0}
\big( \mathsf{K}_l^{(\tau)} \big)_{n, n} &=& - E_W \left(\frac{W}{R}
\right)^2 \frac{K_0^{(n)}}{2} \, \tau \, l \, \eta_0 + E_W \,
\frac{W}{R} \, K_1^{(n)}\, l \, \eta_1 + \frac{1}{2} \left[
E_0^{(\tau,n)} - E_0^{(\tau, -n)} \right] \eta_3 \quad , \\[0.2cm]
\label{eq:aziHam1}
\big( \mathsf{L}_l^{(\tau)} \big)_{n, n} &=& \left\{ \frac{1}{2}
\left[ E_0^{(\tau,n)} + E_0^{(\tau,-n)}\right] + E_W \left(
\frac{W}{R} \right)^2 \left[ \frac{\xi\, L_{0\text{A}}^{(n)}}{k_\Delta W}
\, l^2 - \frac{L_{0\text{B}}^{(n)}}{k_\Delta W} \, \tau\, l \right]\right\}
\eta_0 \nonumber \\[0.2cm]
&& \hspace{8cm} +\, E_W \left( \frac{W}{R} \right)^2 \left\{
\frac{L_{3\text{A}}^{(n)}}{k_\Delta W} \, l^2 - \frac{L_{3\text{B}}^{(n)}}{2}
\, \tau\, l \right\}  \eta_3 \quad . \quad
\end{eqnarray}
\end{subequations}
\end{widetext}
Here $\eta_j$ are Pauli matrices acting in the $2\times 2$
subspace where $\ket{\Phi_0^{(\tau, \pm n)}}$ are the basis states,
i.e., these states correspond to the eigenstates of $\eta_3$ with
eigenvalue $\pm 1$. The dimensionless quantities $K_j^{(n)}$ and
$L_j^{(n)}$ contain relevant parameter dependencies and are given
most generally in terms of matrix elements as
\begin{subequations}\label{eq:dimLessFun}
\begin{eqnarray}
K_0^{(n)} &=& -\tau\frac{\big\langle \sigma_2 \, W/r \big
\rangle^{(\tau)}_{n,n} + \big\langle \sigma_2 \, W/r \big
\rangle^{(\tau)}_{-n,-n}}{(W/R)^2} \quad , \\
K_1^{(n)} &=& \frac{\big\langle \sigma_2 \, W/r \big
\rangle^{(\tau)}_{n,-n}}{W/R} \quad , \\
L_{0\text{A}}^{(n)} &=& \nonumber \\ && \hspace{-0.5cm} \frac{\big\langle
(\sigma_3 + \xi\sigma_0) (W/r)^2 \big\rangle^{(\tau)}_{n,n} + \big
\langle(\sigma_3 + \xi\sigma_0) (W/r)^2 \big\rangle^{(\tau)}_{-n,
-n}}{2\xi (W/R)^2} , \nonumber \\ \\
L_{0\text{B}}^{(n)} &=& \nonumber \\ && \hspace{-0.5cm} \frac{\big\langle
(\sigma_0 + \xi\sigma_3) (W/r)^2 \big\rangle^{(\tau)}_{n,n} + \big
\langle(\sigma_0 + \xi\sigma_3) (W/r)^2 \big\rangle^{(\tau)}_{-n,
-n}}{2 (W/R)^2} , \nonumber \\ \\
L_{3\text{A}}^{(n)} &=& \nonumber \\ && \hspace{-0.5cm} \frac{\big
\langle (\sigma_3 + \xi\sigma_0) (W/r)^2 \big\rangle^{(\tau)}_{n,n}
- \big\langle (\sigma_3 + \xi\sigma_0) (W/r)^2 \big\rangle^{(\tau
)}_{-n,-n}}{2 (W/R)^2} , \nonumber \\ \\
L_{3\text{B}}^{(n)} &=& \tau\frac{\big\langle \sigma_2 \, W/r \big
\rangle^{(\tau)}_{-n,-n} - \big\langle \sigma_2 \, W/r \big
\rangle^{(\tau)}_{n,n}}{(W/R)^2} + \nonumber \\[0.2cm] &&
\hspace{-0.5cm} \frac{\big\langle (\sigma_0 + \xi\sigma_3) (W/r)^2 \big
\rangle^{(\tau)}_{n,n} - \big\langle(\sigma_0 + \xi\sigma_3) (W/r)^2 \big
\rangle^{(\tau)}_{-n, -n}}{(W/R)^2} . \nonumber \\
\end{eqnarray}
\end{subequations}
More explicit expressions for these are available~\cite{yan17} but,
as they are lengthy and unilluminating, we do not present them here.
One important feature is that all the functions given in
Eqs.~(\ref{eq:dimLessFun}a-f) remain finite in the limit $W/R\to 0$,
i.e., the explicit factors $W/R$ in Eqs.~(\ref{eq:aziHam0}) and
(\ref{eq:aziHam1}) constitute the leading behavior in the limit of
narrow rings. The contribution
$\big( \mathsf{L}_l^{(\tau)}\big)_{n, n}$ vanishes in the
ordinary-Dirac case where $\Delta(k)\to\Delta_0$,
$\epsilon(k) \to 0$, i.e., $k_\Delta W\to\infty$ and $\xi\to 0$. In
the same limit, and assuming also a small ring aspect ratio
$W/R\to 0$, we obtain
\begin{subequations}
\begin{eqnarray}\label{eq:univUps} 
K_0^{(n)} &\to& \frac{1 + \Delta_0/(2 E_W)}{\big[ E_0^{(\tau,n)}/E_W
\big]^2 + \Delta_0/(2 E_W)}\quad , \\[0.2cm] K_1^{(n)} &\to& 1 \quad .
\end{eqnarray}
\end{subequations}

\begin{figure}[t]
\includegraphics[width=0.9\columnwidth]{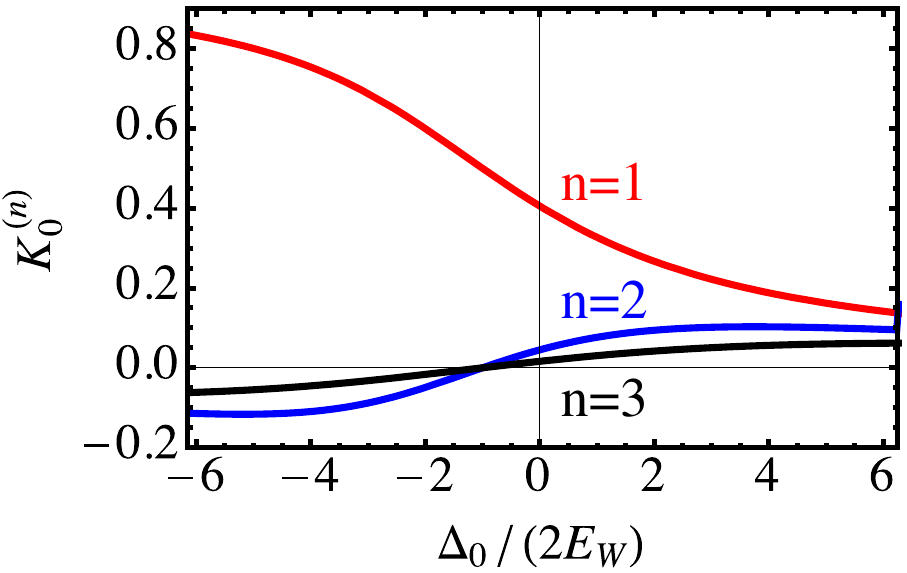}
\caption{\label{fig:SLG_Ups}
Universal system-parameter dependence of the quantity $K_0^{(n)}$
for the ordinary-Dirac case [$\Delta(k)\to \Delta_0$, $\epsilon(k)
\to 0$] in the narrow-ring limit ($W/R\to 0$) according to
Eq.~(\ref{eq:univUps}). The red (blue, black) curve shows the result
for $n=1$ ($2$, $3$). Note the qualitatively different behavior of
the $n=1$ subband in the inverted regime where $\Delta_0/2 < - E_W$.
All curves for $K_0^{(n)}$ with $n>1$ cross at the point $(-1, 0)$ and
exhibit an associated sign change.}
\end{figure}

\noindent
The universal dependence of the quantity $K_0^{(n)}$ on system
parameters in this limit is plotted for the three lowest
positive-energy subbands in Fig.~\ref{fig:SLG_Ups}, revealing a
qualitatively different behavior of the $n=1$ subband. In particular,
$K_0^{(n)}$ for the higher subbands (i.e., for $n>1$) vanishes at the
point $\Delta_0/2 = -E_W$ where the transition between normal and
topological ring-subband structure occurs. The contrasting behavior of
$K_0^{(1)}$ is manifested in its opposite monotonicity and absence of
any sign change. Understanding the behavior of $K_0^{(n)}$ is relevant
because, as discussed in greater detail in Sec.~\ref{sec:transport},
this quantity determines the confinement-induced geometric phase that
features prominently in the quantum-interference contribution to the
Dirac-ring conductance.

The expression given in Eq.~(\ref{eq:aziHam0}) constitutes the
minimal complete model Hamiltonian governing azimuthal motion in a
Dirac-electron quantum-ring subband. It becomes accurate in the
limit of small electron-hole asymmetry $\xi$, yielding the
approximate subband dispersions
\begin{eqnarray}\label{eq:modelDisp}
E_l^{(\tau, \pm n)} &\approx& E_W \Bigg\{ -\left( \frac{W}{R} 
\right)^2 \frac{K_0^{(n)}}{2} \, \tau \, l \nonumber \\
&& \hspace{0.3cm} \pm\, \sqrt{\left( \frac{E_0^{(\tau,n)}}{E_W}
\right)^2 + \left(\frac{W}{R}\right)^2 {K_1^{(n)}}^2 l^2} \Bigg\}
\, . \quad
\end{eqnarray}
The terms $\propto \eta_1$ and $\propto \eta_3$ from
Eq.~(\ref{eq:aziHam0}) are the most familiar~\cite{zar10, sti13,
dco14, bol14}, as they constitute the expected one-dimensional Dirac
form. In particular, the ring confinement induces an effective-gap
contribution $\propto\eta_3$ that is generally the largest term, even
if the 2D material that hosts the Dirac-ring structure has a vanishing
band gap (as is the case, e.g., for graphene). The only possible
exception is the lowest ($n=1$) subband deep in the inverted regime
when $\Delta_0/2 \ll -E_W$, as then $E_0^{(\tau,1)} - E_0^{(\tau, -1)}
\to 0$. See Figs.~\ref{fig:BHZsubbs} and \ref{fig:SLGuniv} for an
illustration. The fact that the size-quantization energy appears like
a mass gap in the ring subband energies was implicit in thorough
treatments of graphene rings~\cite{rec07, bel16} but has sometimes
been overlooked in simplified models~\cite{zar10, sti13, dco14,
bol14}. The contribution $\propto\eta_0$ embodies the breaking of
flavor symmetry due to the ring confinement. For the case of graphene,
where the flavor degree of freedom corresponds to electrons from the
different valleys $\tau\vek{K}$, this was discussed in
Refs.~\onlinecite{rec07, wur10}. Note that, although the term $\propto
\eta_0$ is nominally higher-order in $W/R$ than the term $\propto
\eta_1$, both terms contribute at the same order (quadratic in $W/R$)
to the energy dispersions [cf.\ Eq.~(\ref{eq:modelDisp})] for finite
$E_0^{(\tau,n)}-E_0^{(\tau,-n)}$. 

\begin{figure}[t]
\includegraphics[width=0.9\columnwidth]{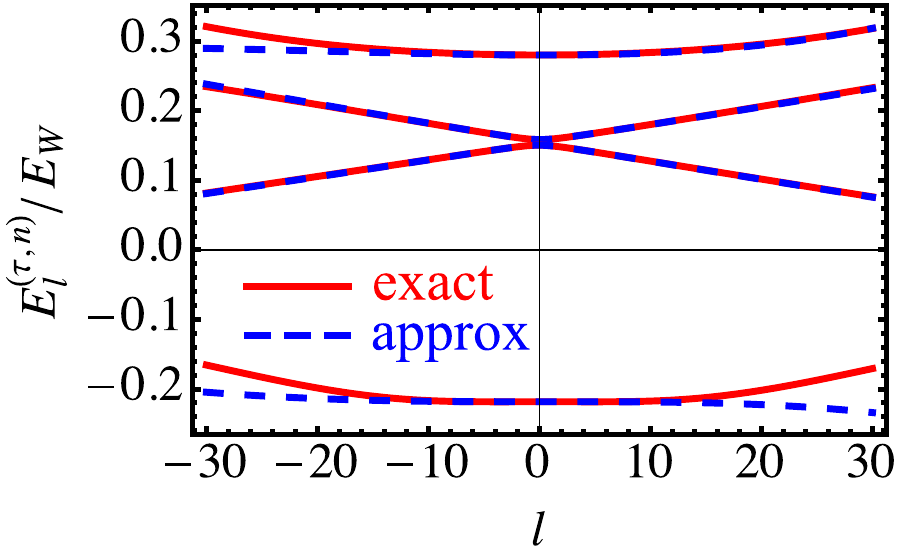}
\caption{\label{fig:offDiagCheck} Comparison of energy dispersions for
subbands with $n=\pm 1, \pm 2$ and $\tau = +$ derived from effective
Hamiltonians Eq.~(\ref{eq:aziHam}) for individual subbands (dashed
blue curves) and the exact spectrum obtained by diagonalizing the full
Hamiltonian matrix Eq.~(\ref{eq:xLham}) (solid red curves). Results
shown were calculated for a ring structure satisfying $k_\Delta W =
26.6$ and $W/R = 0.1$, using band-structure parameters
for a 7-nm HgTe quantum well given in Ref.~\onlinecite{rot10}.}
\end{figure}

The contributions collected in $\big( \mathsf{L}_l^{(\tau)} \big)_{n,
n}$ [cf.\ Eq.~(\ref{eq:aziHam1})] are sub-leading in the sense that
they are proportional to the electron-hole asymmetry $\xi$ or
suppressed by the typically small factor $1/(k_\Delta W)$. However,
in particular in systems with sizable electron-hole asymmetry as,
e.g., HgTe quantum wells, these contributions can become important
enough to necessitate their inclusion. In contrast, the coupling
between subspaces with different $|n|$ turns out to only marginally
affect the low-lying subband dispersions for realistic sets of
materials parameters.  Figure~\ref{fig:offDiagCheck} illustrates the
high level of accuracy typically obtained by using only the effective
Hamiltonian of Eq.~(\ref{eq:aziHam}) to describe the azimuthal
motion of Dirac electrons in the ring structure. Exact and
approximate dispersions pertaining to the lowest pair of subbands
are seen to be virtually indistinguishable, while deviations become
visible for the higher ring subbands.

\section{Conductance and geometric phase for Dirac-electron rings}
\label{sec:transport}

\begin{figure}[t]
\includegraphics[width=0.6\columnwidth]{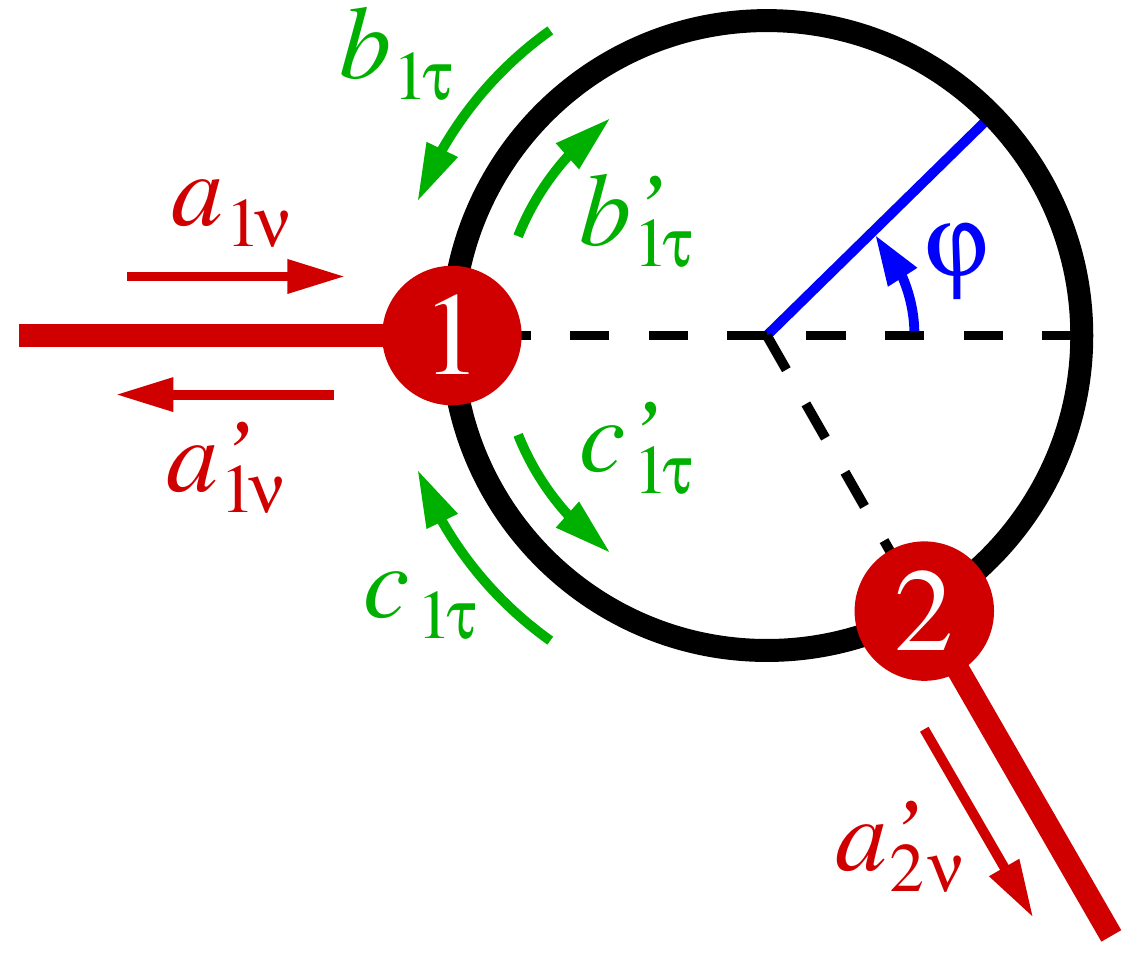}
\caption{\label{fig:ringscatt}
Electronic transport through a quantum-ring conductor. Leads are
connected to the ring at two T junctions from which charge carriers
are incoming (outgoing) with amplitude $a_{j\nu}$ ($a^\prime_{j\nu}$)
in lead mode $\nu$. Propagation in the ring occurs via
confined-Dirac-electron modes having fixed subband index $n$ and
flavor $\tau=\pm$. We assume a clean ring, i.e., no scattering to
occur between modes except at the junctions.}
\end{figure}

To describe electric transport through a quantum ring conductor
realized using a 2D-Dirac material, we consider a situation depicted
schematically in Fig.~\ref{fig:ringscatt}. Electrons are transmitted
from a source lead 1 into a drain lead 2 via the Dirac-ring
eigenstates. The coupling of lead modes $\nu$ to the eigenmodes of
electrons in the ring occurs at the T junctions. To keep the
discussion simple, we assume in the following that only the ring
subband with label $n$ contributes to transport.\footnote{This
approximation strictly applies only at low temperatures when the Fermi
energy in the leads crosses only one of the lowest ($n=\pm 1$) ring
subbands. The generalization to multi-mode configurations that are
more realistic when also higher-$|n|$ subbands are populated can be
obtained using the approach of Ref.~\onlinecite{bue85}.} The linear
electric conductance $G$ is determined by the transmission functions
$T_{\nu_2 \leftarrow \nu_1}(E)$ between source-lead and drain-lead
modes via~\cite{dat95}
\begin{equation}
G = G_0 \, \sum_{\nu_1, \nu_2} T_{\nu_2 \leftarrow \nu_1}
(E_\mathrm{L}) \quad .
\end{equation}
Here $G_0 \equiv g\, e^2/(2\pi\hbar)$ is the universal quantum of
conductance multiplied by a positive integer $g$ counting the
degeneracy associated with degrees of freedom that do not affect
charge-carrier dynamics and are therefore not included in the model
Hamiltonian Eq.~(\ref{eq:4x4Ham}),\footnote{For example, the real
spin of electrons in graphene is accounted for by $g=2$.} and
$E_\mathrm{L}$ is the chemical potential in the leads.

The transmission function depends sensitively on details of the device
structure, especially on how the leads are coupled to the ring. Two
basic physical scenarios can be distinguished according to whether the
ring is (i)~attached to leads that are made of the same material as
the ring, or (ii)~attached to a different material (generally via
tunneling). Case (i) is more typical nowadays, as it is common to
fabricate an entire mesoscopic-conductor system out of a host material
using one of many available lithography techniques~\cite{mad02}. In
that case, the same two-flavor 2D-Dirac dynamics governs
charge-carrier motion in the leads as well as the ring. On the other
hand, contact can also be made to a mesoscopic ring structure using
scanning-probe tips or other nanoelectronic connections, in which case
the charge-carrier dynamics in the leads can be very different from
that in the ring. Such case-(ii) scenarios can also be described
straightforwardly using the scattering approach to quantum transport.
However, as case (ii) is rather uncommon for quantum-ring samples, we
consider here only case (i). In particular, given that the general
goal in experiments is to make good contacts between the ring and the
leads, we assume junctions between the ring and the attached leads to
be sufficiently adiabatic so that the flavor ($\tau$) degree of
freedom is conserved. This general situation also lends itself to
exploring opportunities for \emph{flavortronics\/}, i.e.,
quantum-transport effects that capitalize on $\tau$-dependent
charge-carrier dynamics. In the context of 2D atomic crystals where
$\tau$ corresponds to the valley degree of freedom, this concept is
generally refered to as \emph{valleytronics\/}~\cite{sch16}.

In our formalism, we allow for the possibility of asymmetric ring
structures where the two T junctions with external leads are not
identical and/or are not placed diametrically opposite from each
other. The ring segments connecting them are assumed to be
sufficiently clean so that scattering of charge carriers only occurs
at the T junctions. This is a physically realistic assumption as
recently fabricated mesoscopic structures of 2D-Dirac materials are
ballistic~\cite{kon06, dau17}.

We proceed by presenting the derivation of the transmission functions
in Sec.~\ref{sec:Tjunc}. Our results can be applied to identify
features in the conductance that provide direct measures for the
peculiar electronic properties of Dirac-electron rings. This is
illustrated in greater detail in Sec.~\ref{sec:condRes}, together with
implications for using quantum ring conductors as flavor-filtering
devices.

\subsection{General transmission function for
\texorpdfstring{\bm{$\tau$}}{}-conserving ballistic ring structures}
\label{sec:Tjunc}

The procedure for determining the transmission functions $T_{\nu_2
\leftarrow\nu_1}(E)$ through a ring conductor is based on two
fundamental ingredients~\cite{gef84, bue84}. Firstly, the coupling
between lead states and ring states at fixed energy $E$ is embodied by
the \textit{S} matrix~\cite{dat95} of each T junction~\cite{sha83, 
ito95}. Secondly, because we assume no scattering to occur in the ring
segments connecting the leads, the quantum amplitudes of ring states
at different junctions are related simply by the dynamical phases
corresponding to propagation of the ring eigenstates between them.
These relationships enable the algebraic elimination of ring-state
amplitudes, yielding an expression for the outgoing lead-2 amplitudes
$a_{2\nu_2}^\prime$ in terms of incoming lead-1 amplitudes $a_{1
\nu_1}$ and thus the transmission functions $T_{\nu_2 \leftarrow
\nu_1}(E)\equiv |a_{2\nu_2}^\prime/ a_{1\nu_1}|^2$.

In our situation of interest, charge carriers having different flavor
$\tau$ are transmitted through the combined leads-and-ring structure
completely in parallel. Scattering at the T junctions then occurs only
between modes with the same $\tau$,
\begin{equation}
\left( \begin{array}{c} a_{j\tau}^\prime \\[0.1cm] b_{j\tau}^\prime
\\[0.1cm] c_{j\tau}^\prime \end{array} \right) = \,
\underline{S}_{j\tau} \left( \begin{array}{c} a_{j\tau} \\[0.1cm]
b_{j\tau} \\[0.1cm] c_{j\tau} \end{array}\right)\quad ,
\end{equation}
and we adopt the most general form for the \textit{S} matrices,
\begin{equation}\label{eq:genSmat}
\underline{S}_{j\tau} = \left( \begin{array}{ccc} -\sqrt{1-2
\varepsilon_{j\tau}}\,\ee^{i \psi_{j\tau}} & \sqrt{\varepsilon_{j
\tau}} & \sqrt{\varepsilon_{j\tau}} \\[0.2cm] \sqrt{\varepsilon_{j
\tau}} & \kappa_{j\tau}\, \ee^{-i \psi_{j\tau}} & \lambda_{j\tau}\,
\ee^{-i \psi_{j\tau}}\\[0.2cm] \sqrt{\varepsilon_{j\tau}} &
\lambda_{j\tau}\, \ee^{-i\psi_{j\tau}} & \kappa_{j\tau} \, \ee^{-i
\psi_{j\tau}} \end{array} \right) .
\end{equation}
Here the parameters $\varepsilon_{j\tau}$ with $0\le \varepsilon_{j
\tau} \le 1/2$ are a measure for how strongly lead $j$ is coupled to
the ring via mode $\tau$, with $\varepsilon_{j\tau}=1/2$
($\varepsilon_{j\tau}=0$) describing the extremal situation of a fully
transparent junction (a completely isolated ring). Scattering of
$\tau$-flavor electrons between the ring segments at junction $j$ is
described by reflection amplitudes
\begin{subequations}
\begin{equation}
\kappa_{j\tau} = |\kappa_{j \tau}|\, \ee^{i (\phi_{j\tau} +
\varrho_{j\tau})}
\end{equation}
and transmission amplitudes
\begin{equation}
\lambda_{j\tau} = |\lambda_{j\tau}|\, \ee^{i(\phi_{j\tau} -
\varrho_{j\tau})} \quad .
\end{equation}
\end{subequations}
The real but otherwise arbitrary phases $\psi_{j\tau}$ are associated
with back-reflection into the leads. The canonical expression for
$\underline{S}_{j\tau}$ given in Eq.~(\ref{eq:genSmat}) covers
previously considered special cases of purely real~\cite{sha83, bue84}
or symmetric-beam-splitter~\cite{gia11} T-junction \textit{S}
matrices, as well as the general form given in Ref.~\onlinecite{kow90}.
Unitarity of $\underline{S}_{j\tau}$ imposes the relations
\begin{subequations}
\begin{eqnarray}
1 &=& |\kappa_{j\tau}|^2 + |\lambda_{j\tau}|^2 + \varepsilon_{j\tau}
\quad , \\[0.2cm] \varrho_{j\tau} &=& \frac{s_{j\tau}}{2}\, \arccos
\left(\frac{-\varepsilon_{j\tau}}{2 |\kappa_{j\tau}|\, |\lambda_{j
\tau}|} \right) \quad , \\[0.2cm] \phi_{j\tau} &=& \arctan \left[
\left|\frac{|\lambda_{j\tau}| - |\kappa_{j\tau}|}{|\lambda_{j\tau}|
+ |\kappa_{j\tau}|} \right| \tan\varrho_{j\tau}\right] \quad ,
\end{eqnarray}
with
\begin{equation}
s_{j\tau} = \left\{ \begin{array}{cl} \mathrm{sgn}(|\lambda_{j\tau}|
-|\kappa_{j\tau}|) & \mbox{ if } |\kappa_{j\tau}| \ne |\lambda_{j
\tau}| \,\, , \\[0.2cm] \pm 1& \mbox{ otherwise} \,\, . \end{array}
\right. 
\end{equation}
\end{subequations}

The relation between quantum amplitudes of ring states at different
junctions can be found from the general form of a ring state $\ket{j
\tau; \varphi}$ in mode $\tau$ emanating from junction $j$, which is
a superposition of counterclockwise-moving and clockwise-moving
Dirac-ring eigenstates $\ket{j \tau;\varphi}_\pm$,
\begin{subequations}
\begin{eqnarray}\label{eq:juncState}
\ket{j \tau; \varphi} &=& \ket{j \tau; \varphi}_+ + \ket{j \tau;
\varphi}_- \quad , \\[0.2cm]
\ket{j \tau; \varphi}_+ &=& c_{j \tau}^\prime \,\, \ee^{i l_+^{(\tau)}
(\varphi - \varphi_j)}\,\, \frac{\mathcal{U}_\tau (\varphi -
\varphi_j)}{\sqrt{2 \pi r}}\, \ket{\Phi_{l_+^{(\tau)}}^{(\tau, n)}}
\, , \\[0.2cm] \label{eq:clockwise}
\ket{j \tau; \varphi}_- &=&  - b_{j \tau}^\prime \,\, \ee^{i
l_-^{(\tau)}(\varphi - \varphi_j - 2\pi)}\,\, \frac{\mathcal{U}_\tau
(\varphi - \varphi_j)}{\sqrt{2\pi r}} \,
\ket{\Phi_{l^{(\tau)}_-}^{(\tau, n)}} \, . \nonumber \\
\end{eqnarray}
\end{subequations}
Here $\varphi_j$ denotes the location of lead $j$, with the
conventions $-\pi \le \varphi_j < \pi$ and $\varphi > \varphi_j$. As
shown in Fig.~\ref{fig:ringscatt}, the counterclockwise (clockwise)
outgoing mode at junction $j$ propagating in channel $\tau$ has
amplitude $c_{j\tau}^\prime$ ($b_{j\tau}^\prime$). Due to the ring
geometry, the azimuthal angle $\varphi$ for clockwise-moving partial
waves acquires a phase shift $2\pi$ with respect to that for
counterclockwise-moving partial waves. We used the relation
$\mathcal{U}_\tau(\varphi - \varphi_j - 2\pi) = -\mathcal{U}_\tau
(\varphi -\varphi_j)$ in Eq.~(\ref{eq:clockwise}). Given $E$, the
azimuthal quantum numbers $l^{(\tau)}_+$ and $l^{(\tau)}_-$ are
determined from the relation $E = E_{l^{(\tau)}_\pm}^{(\tau, n)}$
and, by definition, $d E_{l}^{(\tau,n)}/d l\big|_{l=l^{(\tau)}_\pm}
\gtrless 0$. For the 2D-Dirac-material ring conductors considered
here, we have $l^{(\tau)}_+ \ne - l^{(\tau)}_-$ in general, but
time-reversal symmetry mandates~\cite{rec07}
\begin{equation}\label{eq:pmMomenta}
l^{(\tau)}_\pm = - l^{(-\tau)}_\mp \quad .
\end{equation}
Assuming no scattering to occur within the ring segments between the
junctions and considering the situation with $\varphi_2 > \varphi_1$,
the form of $\ket{1 \tau; \varphi_2}$ ($\ket{2 \tau; \varphi_1 +
2\pi}$) determines how the incoming amplitudes at junction 2 (1)
depend on the outgoing amplitudes of junction 1 (2) on the same
segment. Using a compact transfer-matrix notation, we can write
\begin{subequations}
\begin{eqnarray}
\left( \begin{array}{c} c_{2\tau} \\ c_{2\tau}^\prime \end{array}
\right) &=& \ee^{i( \theta_\tau - \bar\theta_\tau - \pi)} \left(
\begin{array}{cc} \ee^{-i ( \chi - \bar\chi )} & 0 \\ 0 & \ee^{i (
\chi - \bar\chi )} \end{array} \right) \left( \begin{array}{c}
b_{1\tau}^\prime \\ b_{1\tau}\end{array}\right) , \nonumber\\
\\[0.2cm] \left( \begin{array}{c} c_{1\tau} \\ c_{1\tau}^\prime
\end{array}\right) &=& \ee^{-i( \theta_\tau + \bar\theta_\tau)}
\left( \begin{array}{cc} \ee^{i ( \chi + \bar\chi )} & 0 \\ 0 &
\ee^{-i ( \chi + \bar\chi )} \end{array} \right) \left(
\begin{array}{c} b_{2\tau}^\prime \\ b_{2\tau}\end{array}\right) ,
\nonumber \\
\end{eqnarray}
\end{subequations}
in terms of phases
\begin{subequations}
\begin{eqnarray}
\theta_\tau &=& \frac{1}{2} \left[ l_+^{(\tau)} + l_-^{(\tau)}\right]
\left( \varphi_2 - \varphi_1 - \pi \right) \quad , \\[0.2cm]
\chi &=& \frac{1}{2} \left[ l_+^{(\tau)} - l_-^{(\tau)} \right] \left(
\varphi_2 - \varphi_1 - \pi \right)
\end{eqnarray}
\end{subequations}
that depend on the T-junction locations, and the purely
electronic-structure-determined phases
\begin{subequations}
\begin{eqnarray}\label{eq:barTheta}
\bar\theta_\tau &=& \frac{\pi}{2} \left[ l_+^{(\tau)} + l_-^{(\tau)}
\right] \quad , \\[0.2cm] \label{eq:barChi}
\bar\chi &=& \frac{\pi}{2} \left[ l_+^{(\tau)} - l_-^{(\tau)} \right]
\end{eqnarray}
\end{subequations}
that are intrinsic measures of interference in the Dirac ring. By
construction, $\theta_\tau$ and $\chi$ vanish for a symmetric ring
structure where $\varphi_2-\varphi_1\equiv\pi$, and
Eq.~(\ref{eq:pmMomenta}) implies that $\chi$ and $\bar\chi$ do not
depend on $\tau$ whereas $\bar\theta_\tau\equiv \tau\, \bar\theta_+$.

The transmission function for the situation where scattering at the
junctions conserves $\tau$ can be written as $T_{\nu_2\leftarrow\nu_1}
\equiv T_\tau\, \delta_{\nu_1 \tau} \delta_{\nu_2 \tau}$, with $T_\tau
= |a^\prime_{2\tau}/a_{1\tau}|^2$. A straightforward calculation
yields the fully general result
\begin{widetext}
\begin{equation}\label{eq:tauTrans}
T_\tau \big( \chi, \tilde\chi_\tau, \theta_\mathrm{AA}^{(\tau)} 
\big) = \frac{4\varepsilon_{1\tau} \varepsilon_{2\tau} \left[ \cos^2 
\chi\, \sin^2 \tilde\chi_\tau\, \cos^2\big(\theta_\mathrm{AA}^{(\tau)}
/2\big) + \sin^2\chi\, \cos^2\tilde\chi_\tau\, \sin^2\big(
\theta_\mathrm{AA}^{(\tau)}/2\big) \right]}{\left| |\kappa_{1\tau}| 
|\kappa_{2\tau}| \ee^{i (\varrho_{1\tau} + \varrho_{2\tau})} 
\cos(2\chi) + |\lambda_{1\tau}| |\lambda_{2\tau}|\ee^{-i(\varrho_{1
\tau} + \varrho_{2\tau})}\cos\theta_\mathrm{AA}^{(\tau)} -F_\tau(
\omega_{1\tau} + \omega_{2\tau}, \phi_{1\tau} + \phi_{2\tau},
2\tilde\chi_\tau) \right|^2} \,\, ,
\end{equation}
where
\begin{subequations}
\begin{equation}
F_\tau(\omega, \phi, 2\tilde\chi) = \frac{1}{2} \left[\ee^{i
\omega} + \sqrt{(1 - 2\varepsilon_{1\tau}) (1-2\varepsilon_{2\tau})}
\, \ee^{-i\phi}\right]\cos(2\tilde\chi) - \frac{i}{2} \left[ \ee^{i
\omega} -\sqrt{(1 - 2 \varepsilon_{1\tau})(1-2\varepsilon_{2\tau})}
\, \ee^{-i\phi} \right]\sin(2\tilde\chi) \, ,
\end{equation}
\end{subequations}
\end{widetext}
and
\addtocounter{equation}{-1}
\begin{subequations}
\addtocounter{equation}{1}
\begin{equation}\label{eq:AAphase}
\theta_\mathrm{AA}^{(\tau)} \equiv 2\bar\theta_\tau + \pi
\end{equation}
is the generalized Berry~\cite{ber84}, or
Aharonov-Anandan~\cite{aha87}, phase for an individual
$\tau$-conserving transport channel in the Dirac ring. We also used
the abbreviations
\begin{eqnarray}\label{eq:nonunivChi}
\tilde\chi_\tau &=& \bar\chi - \frac{1}{2}\left(\psi_{1\tau} + \psi_{2
\tau} - \phi_{1\tau} - \phi_{2\tau} - \omega_{1\tau} - \omega_{2\tau}
\right)\,\, , \nonumber \\ \\[0.2cm]
\omega_{j\tau} &=& \arg(\lambda_{j\tau} - \kappa_{j\tau}) - \phi_{j
\tau} \quad , \\[0.2cm] &\equiv&  \arctan \left[ \left| \frac{|
\lambda_{j\tau}| + |\kappa_{j\tau}|}{|\lambda_{j\tau}| - |\kappa_{j
\tau}|} \right| \tan\varrho_{j\tau}\right] \quad .
\end{eqnarray}
\end{subequations}

Among the interesting insights that can be gleaned from
Eq.~(\ref{eq:tauTrans}) are, firstly, that the phase $\theta_\tau$
does not appear at all in the expression for $T_\tau$ and, secondly,
that nonuniversal scattering phases due to the coupling to leads only
enter via $\tilde\chi_\tau$. In the limit $\chi = 0$ and $\varrho_{j
\tau} = \pi/2$, our result Eq.~(\ref{eq:tauTrans}) has the form found
previously in Ref.~\onlinecite{bue84} for the transmission through a
symmetric quantum-ring geometry with real \textit{S} matrices used
to describe the T junctions, and the dependence on $\chi$ is
consistent with previously considered cases~\cite{fol05, aeb05} of
quantum rings with arbitrary location of lead-attachment points.

\subsection{Dirac-ring conductance and \texorpdfstring{\bm{$\tau$}}{}
filtering}\label{sec:condRes}

For our case of interest where the $\tau$ degree of freedom is
conserved, we can construct two transport-related quantities of
interest. One is the total electric conductance,
\begin{equation}
G = G_0\sum_{\tau = \pm} T_\tau \quad ,
\end{equation}
and the other is the $\tau$-polarization of the conductance,
\begin{equation}
P_\tau = \frac{T_\tau - T_{-\tau}}{T_\tau + T_{-\tau}} \quad .
\end{equation}
In principle, the fully general Eq.~(\ref{eq:tauTrans}) for the
transmission function $T_\tau$ contains all possible $\tau$-dependent
effects arising from the coupling to the leads via T junctions, as
well as those due to the special features of the Dirac-ring subband
structure. In the following, we will focus on discussing the latter
and therefore assume that the T-junction parameters are the same for
both $\tau=\pm$. To further simplify the discussion, we will consider
the case of leads being attached exactly opposite each other
($\varphi_2-\varphi_1=\pi$), i.e., $\chi\equiv 0$. Thus the only
remaining $\tau$-dependent quantity is the phase $\bar\theta_\tau$,
and we set $\tilde\chi_\tau\equiv\tilde\chi$ from now on. It is then
instructive to apply the approximate expression
Eq.~(\ref{eq:modelDisp}) for the $n$th ring-subband dispersion to
determine $l^{(\tau)}_\pm$ and, via Eqs.~(\ref{eq:barTheta}) and
(\ref{eq:barChi}), the phases $\bar\theta_\tau$ and $\bar\chi$.
Considering the case of a narrow ring ($W/R\ll 1$) and $E_\mathrm{L}
- E_0^{(\tau, n)}\ll E_0^{(\tau, n)}$, we find
\begin{subequations}
\begin{eqnarray}\label{eq:SLGphase}
\bar\theta_\tau &=& \tau\,\frac{\pi}{2}\,\frac{E_0^{(\tau, n)}}{E_W}
\, \frac{K_0^{(n)}}{\big(K_1^{(n)}\big)^2} \quad , \\[0.1cm]
\bar\chi &=& \frac{\pi}{K_1^{(n)}}\, \sqrt{\frac{2 E_0^{(\tau,
n)}}{E_W}} \,\sqrt{\frac{E_\mathrm{L} - E_0^{(\tau, n)}}{E_W (W/R)^2}}
\quad .
\end{eqnarray}
\end{subequations}

\begin{figure}[b]
\includegraphics[width=0.85\columnwidth]{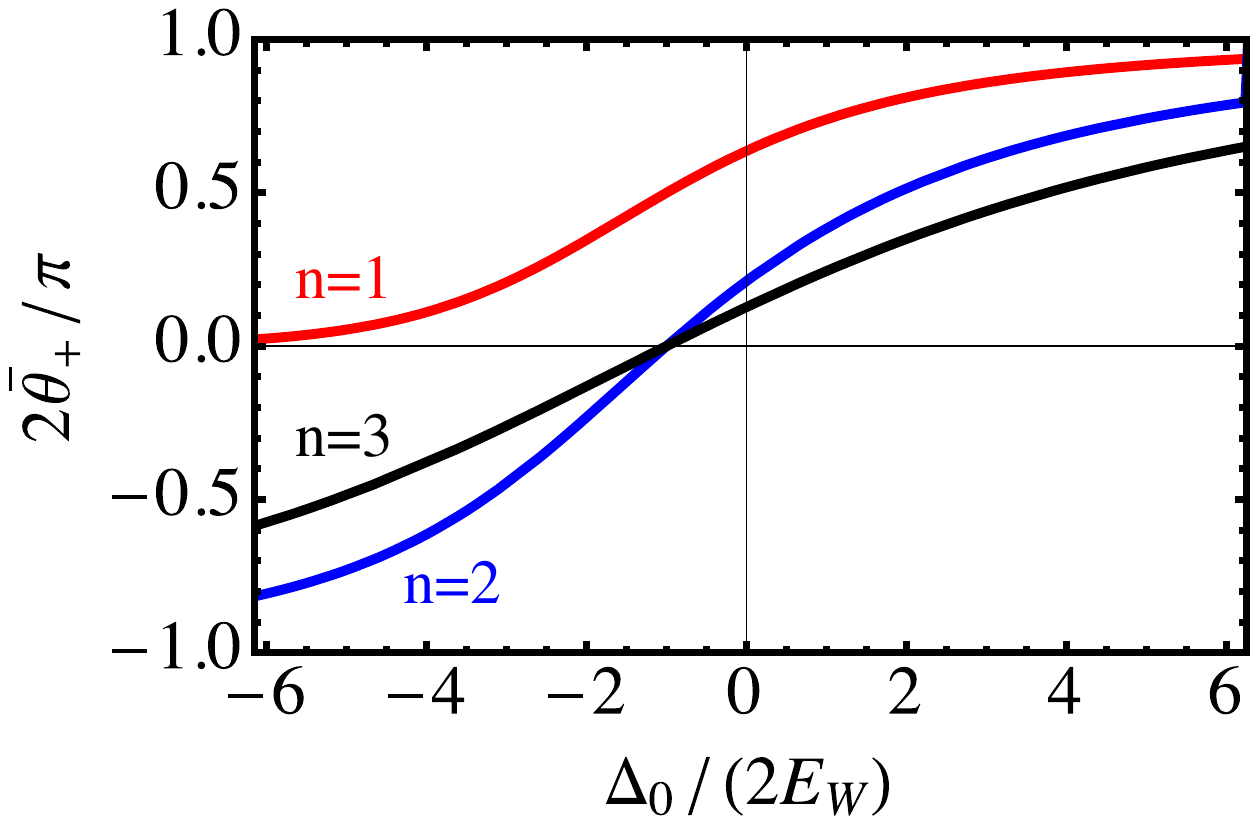}
\caption{\label{fig:SLGphase}%
Electronic-structure-related contribution $2\bar\theta_+$ to a Dirac
ring's intrinsic Aharonov-Anandan phase that is associated with the
$\tau=+$ transport channel. The figure shows this phase for the
subband with $n=1$ ($2$, $3$) as the red (blue, black) curve for a
narrow-ring structure in the ordinary-Dirac limit where $\Delta(k)=
\Delta_0$ and $\epsilon(k)=0$ (corresponding, e.g., to single-layer
graphene). Note the qualitatively different behavior of the $n=1$
subband for which $2\bar\theta_+\approx 0$ deep in the inverted
regime, whereas $|2\bar\theta_+| \approx \pi$ when $|\Delta_0| \gg
E_W$ for the higher subbands.}
\end{figure}

Interestingly, $\bar\theta_\tau\equiv \tau\,\bar\theta_+$ turns out
to be determined solely by the Dirac-ring subband structure. As an
illustration, Fig.~\ref{fig:SLGphase} shows a plot of $2\bar\theta_+$
that has been calculated from Eq.~(\ref{eq:SLGphase}) for the three
lowest subbands in the ordinary 2D-Dirac limit where $\Delta(k)=
\Delta_0$ and $\epsilon(k)=0$. The properties of the curves
representing individual subbands can be traced back directly to the
behavior of the quantities $E_0^{(\tau, n)}$ and $K_0^{(n)}$ for these
subbands that are shown in Figs.~\ref{fig:SLGuniv} and
\ref{fig:SLG_Ups}, respectively. For example, the confinement-induced
geometric phase $\bar\theta_\tau$ vanishes for all subbands with $n>1$
when $\Delta_0/2 = -E_W$ whereas the contribution of the lowest
($n=1$) remains finite. Furthermore, while the higher subbands recover
the ordinary-Schr\"odinger-electron limit $|2 \theta_+| \approx \pi$
when $|\Delta_0| \gg E_W$ both in the normal and topological (i.e.,
inverted-band) regimes, the geometric phase of the lowest subband
approaches the massless-2D-Dirac limit $2\theta_+ = 0$ deep in the
topological regime.

In contrast to $\bar\theta_\tau$, $\bar\chi$ (and thus also $\tilde
\chi_\tau \equiv \tilde\chi$ here) has a strong dependence on the
chemical potential $E_\mathrm{L}$ in the leads. Given that $\tilde
\chi$ is renormalized by nonuniversal phase shifts associated with
scattering at the T junctions [see Eq.~(\ref{eq:nonunivChi})],
features exhibited in the energy dependence of the transmission
function will generally be quite sample-specific --- even if the
T-junction \textit{S}-matrix parameters vary only weakly in the
experimentally relevant range of energy.

In addition to adjusting the chemical potential in the leads,
application of a perpendicular magnetic field $\vek{B}$ can also be
used as an experimentally accessible knob to manipulate transport
through a ring conductor via Aharonov-Bohm interference~\cite{aha59, 
gef84, bue84, rec07}. Formally, the effect of finite $\vek{B}\equiv 
\vek{\nabla}\times\vek{A}$ in a sufficiently narrow ring can be
modeled by introducing an infinitesimally thin tube of magnetic flux
$\psi$ piercing the ring plane at its origin via the vector potential
$\vek A = [\psi/(2\pi r)]\, \vek{\hat\varphi}$. Making the required
substitutions ${\mathcal H}^{(\pm)}(\vek{k}) \to {\mathcal H}^{(\pm)}
(\vek{k} + 2\pi\vek{A}/\psi_0)$ in Eq.~(\ref{eq:4x4Ham}), with
$\psi_0$ denoting the magnetic flux quantum, translates into the
changes $ k_\varphi \to k_\varphi + \psi/(\psi_0 r)$ and ${\mathcal
V}^{(\tau)}_l (r) \to {\mathcal V}^{(\tau)}_{l + \psi/\psi_0}(r)$. Thus the effect of
the magnetic flux is to rigidly shift the quantum-ring dispersions in
$l$ by the amount $\psi/\psi_0$, with the only ramification for the
transmission function Eq.~(\ref{eq:tauTrans}) being that the
Aharonov-Anandan phase becomes flux-dependent,
\begin{equation}\label{eq:fluxAAphase}
\theta_\mathrm{AA}^{(\tau)} \equiv \theta_\mathrm{AA}^{(\tau)}(\psi)
= 2\bar\theta_\tau + 2\pi\left(\frac{1}{2} - \frac{\psi}{\psi_0}
\right)\quad ,
\end{equation}
resulting in a $\psi_0$-periodic modulation of the conductance as a
function of flux $\psi$~\cite{gef84, bue84}. As $\bar\theta_\tau$ is
robust with respect to changes in the leads' chemical potential and
details of the ring-lead junction morphology, it will be possible
to measure its magnitude via the magnetic-field dependence of the
conductance. In particular, it will be interesting to map the
transition from the nonrelativistic, Schr\"odinger-like regime
$\Delta_0 \gg E_W$ (where $2\bar\theta_+ \approx\pi$) to the
ultrarelativistic, Dirac-like regime $-\Delta_0 \gg E_W$ (where $2
\bar\theta_+\approx 0$) exhibited by the $n=1$ subband (cf.\
Fig.~\ref{fig:SLGphase}). In contrast, the higher subbands show
nonrelativistic behavior whenever $|\Delta_0| \gg E_W$, regardless of
the existence of band inversion.

\begin{figure}[b]
\includegraphics[width=0.85\columnwidth]{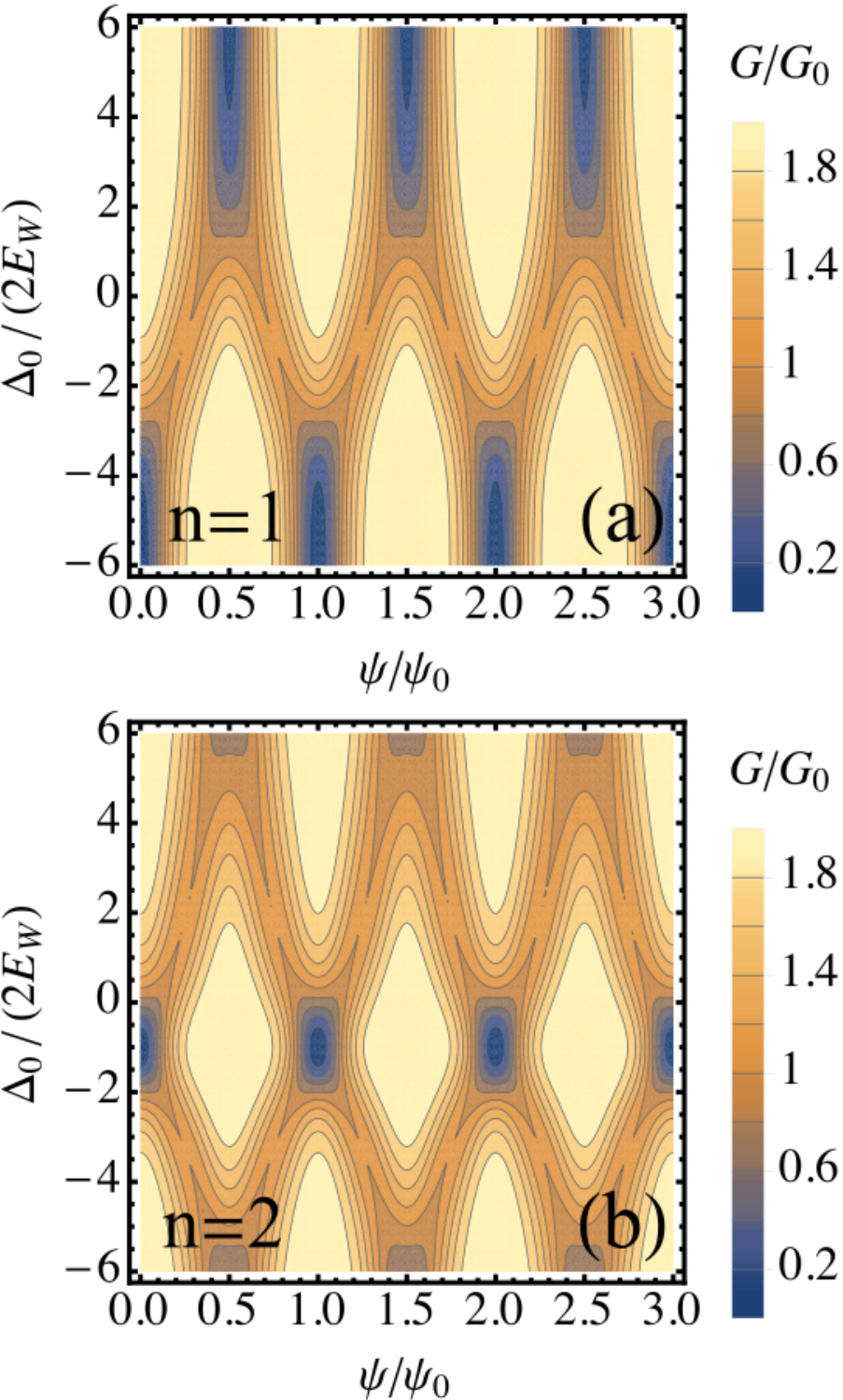}
\caption{\label{fig:CondDens}
Two-terminal conductance $G$ through a ring realized in a material
where charge-carrier dynamics mimics that of ordinary 2D-Dirac
electrons [$\Delta(k)=\Delta_0$, $\epsilon(k)=0$] contacted to leads
via fully transparent T junctions. The density plots show $G$ in units
of $G_0 \equiv g\, e^2/(2\pi\hbar)$ as a function of magnetic flux
$\psi$ in units of the flux quantum $\psi_0\equiv 2\pi \hbar/e$ and
the parameter combination $\Delta_0/(2E_W)$ characterizing the ring
confinement. Results in panel~(a) [(b)] were calculated using
Eq.~(\ref{eq:transpTrans}) for the situation where transport occurs
through the $n=1$ [$n=2$] subband and assuming $\sin^2\tilde\chi =
1$.}
\end{figure}

To illustrate more directly how these distinctive ring-subband
properties are manifested in the two-terminal conductance, we consider
the situation with fully transparent T junctions, which is realized
for $\varepsilon_{j\tau}\equiv 1/2$. As concomitantly $\varrho_{j\tau}
= \pm\pi/2$, $\phi_{j\tau} = 0$, $\omega_{j\tau} = \pm\pi/2$, and
keeping with the current assumption of a symmetric ring structure
where $\chi = 0$ and also $\tilde\chi_{j\tau} \equiv \tilde\chi$,
Eq.~(\ref{eq:tauTrans}) specializes to
\begin{eqnarray}\label{eq:transpTrans}
&& T_\tau^{(\mathrm{tr})} \big( \tilde\chi, \theta_\mathrm{AA}^{(
\tau)}\big) = \nonumber \\ && \hspace{1cm} \frac{2\, \sin^2 \tilde
\chi \, \big( 1 + \cos\theta_\mathrm{AA}^{(\tau)} \big)}{\left(
\frac{1}{2} \big[ 1 + \cos\theta_\mathrm{AA}^{(\tau)}\big] - \cos
(2\tilde\chi)\right)^2 + \sin^2(2\tilde\chi)} \,\, . \quad 
\end{eqnarray}
For the purpose of the present discussion, we fix $\sin^2\tilde\chi=1$
for simplicity. Figure~\ref{fig:CondDens} shows a density
plot of the two-terminal conductance through an ordinary-2D-Dirac ring
that could be realized, e.g., in graphene, as a function of applied
magnetic flux and the quantity $\Delta_0/(2 E_W)$. Results are shown
for two cases corresponding to situations where transport occurs via
states in the lowest ($n=1$) and first excited ($n=2$) ring subband,
respectively. The characteristic dependence of the geometric phase on
ring-structure parameters is clearly exhibited in the
interference-fringe pattern of the conductance. In particular,
massless-Dirac (ordinary-Schr\"odinger) behavior is manifested here
by conductance minima occurring for integer (half-integer) values of
$\psi/\psi_0$. The pattern seen for the $n=1$ subband shows very
clearly a transition between these two limiting regimes. In contrast,
the $n=2$ subband (like all other higher-$|n|$ subbands) exhibits an
interference pattern indicative of massless-Dirac behavior only in a
narrow region around the point where $\Delta_0/2 = -E_W$, which is
a direct consequence of the vanishing $K_0^{(n)}$ for $n>1$ at this
special point where the transition between normal and topological
ring-subband structures occurs (cf.\ Fig.~\ref{fig:SLG_Ups}). For the
subband $n=1$ and $\Delta_0 = 0$, the oscillations of the ring
conductance calculated here as a function of magnetic flux agree with
numerical results presented in Ref.~\onlinecite{wur10} for graphene rings in
the one-mode regime.

\begin{figure}[t]
\includegraphics[width=0.9\columnwidth]{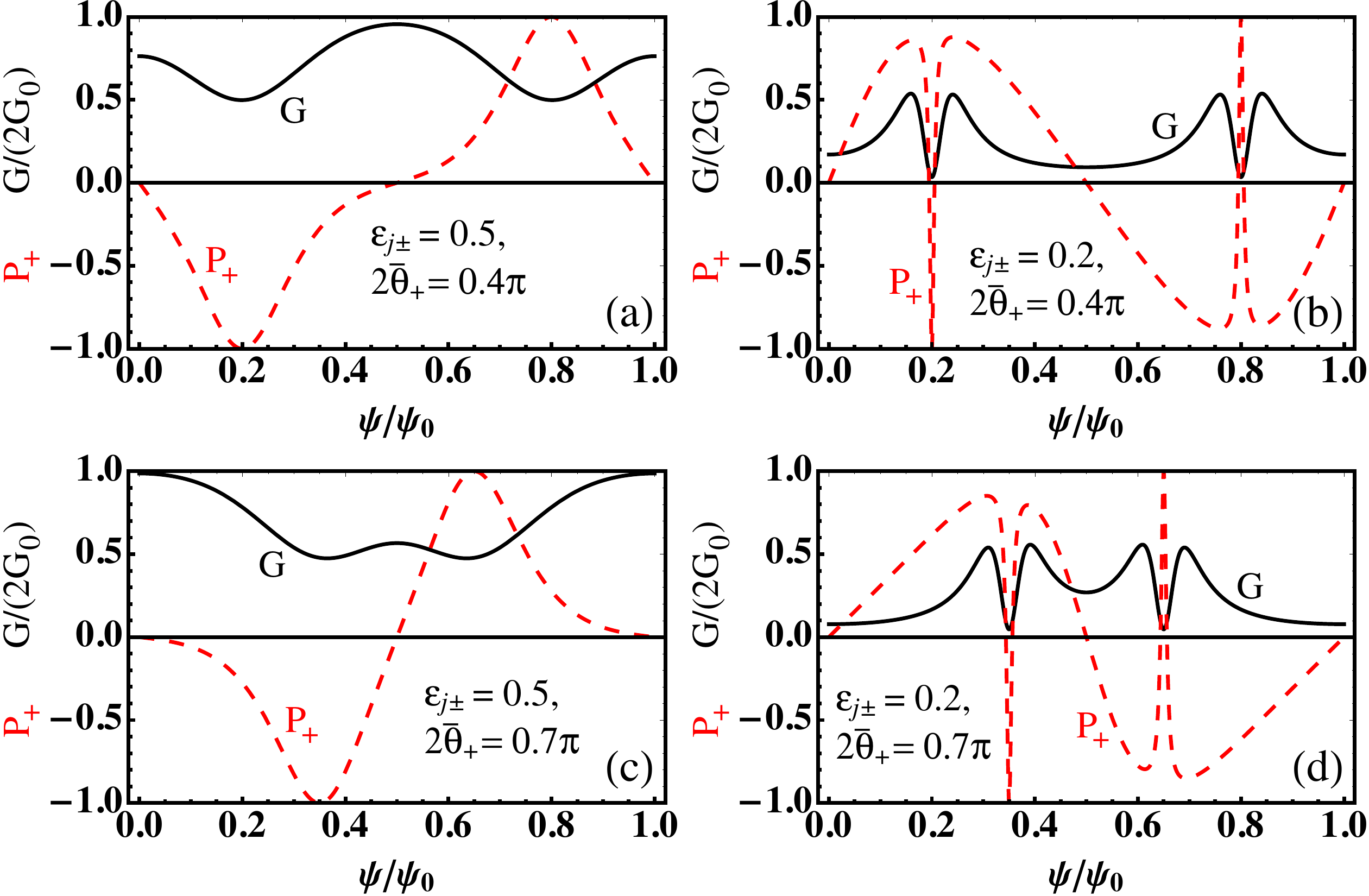}
\caption{\label{fig:realCond}%
Dependence of the conductance $G/(2G_0)$ (black solid curves) and
$\tau$ polarization $P_+$ (red dashed curves) on magnetic flux $\psi$
for a symmetric ring structure where T-junction \textit{S} matrices
are assumed to be real as, e.g., in Ref.~\onlinecite{bue84}. Here $\psi_0$
denotes the quantum of magnetic flux. Other parameters used in the
calculation are $\sin^2\tilde\chi_\pm=1$, $s_{j\pm} = 1$ ($j=1,2$),
and $\varepsilon_{j\pm} = 0.5$ [$0.2$, $0.5$, $0.2$], $2\bar\theta_+
= 0.4\,\pi$ [$0.4\,\pi$, $0.7\, \pi$, $0.7\,\pi$] for panel (a) [(b),
(c), (d)].}
\end{figure}

Before concluding, we explore the dependence of Dirac-ring
interference and flavor filtering on the transparency of the T
junctions connecting the ring to external leads. Two special cases are
considered according to whether T junctions are described by real
\textit{S} matrices as in Ref.~\onlinecite{bue84} or beam-splitter-type
\textit{S} matrices as in Ref.~\onlinecite{gia11}.

Real \textit{S} matrices describing the ring-lead coupling are
obtained from the general expression Eq.~(\ref{eq:genSmat}) by setting
$\varrho_{j\tau} =\pi/2$ and $\psi_{j\tau} =0$ or $\pi$. With the
additional assumptions $\chi=0$ and $\varepsilon_{j\tau}\equiv
\varepsilon$ for a symmetric ring structure, as well as $\sin^2\tilde
\chi_\tau=1$ for simplicity, the transmission function from
Eq.~(\ref{eq:tauTrans}) specializes to
\begin{equation}
T_\tau^\mathrm{(re)}\big (\theta_\mathrm{AA}^{(\tau)}\big) = \frac{2
\varepsilon^2 \big(1 + \cos\theta_\mathrm{AA}^{(\tau)}\big)}{\left[
\frac{(\sqrt{1 - 2 \varepsilon}+1)^2}{4} \big(1 + \cos
\theta_\mathrm{AA}^{(\tau)}\big) + \frac{(\sqrt{1 - 2\varepsilon}
-1)^2}{2} \right]^2} .
\end{equation}
The magnetic-flux dependence of the total Dirac-ring conductance $G$
as well as the flavor ($\tau$) polarization $P_+\equiv -P_-$ for this
situation is illustrated in Fig.~\ref{fig:realCond} for particular
parameter values, including examples for fully transparent T junctions
($\varepsilon=0.5$) and more weakly coupled leads ($\varepsilon=0.2$).
In this case, interference-related minima in $G$ generally coincide
with maxima of $|P_+|$. The flux values at which these features occur
can be shifted by tuning the confinement-related Aharonov-Anandan
angle $\bar\theta_+$. Reduced transparency of the contacts with leads
results in a precipitous narrowing of their flux-dependence line shape
into resonances that are indicative of the isolated-ring bound-state
energies~\cite{bue84}.

\begin{figure}[t]
\includegraphics[width=0.9\columnwidth]{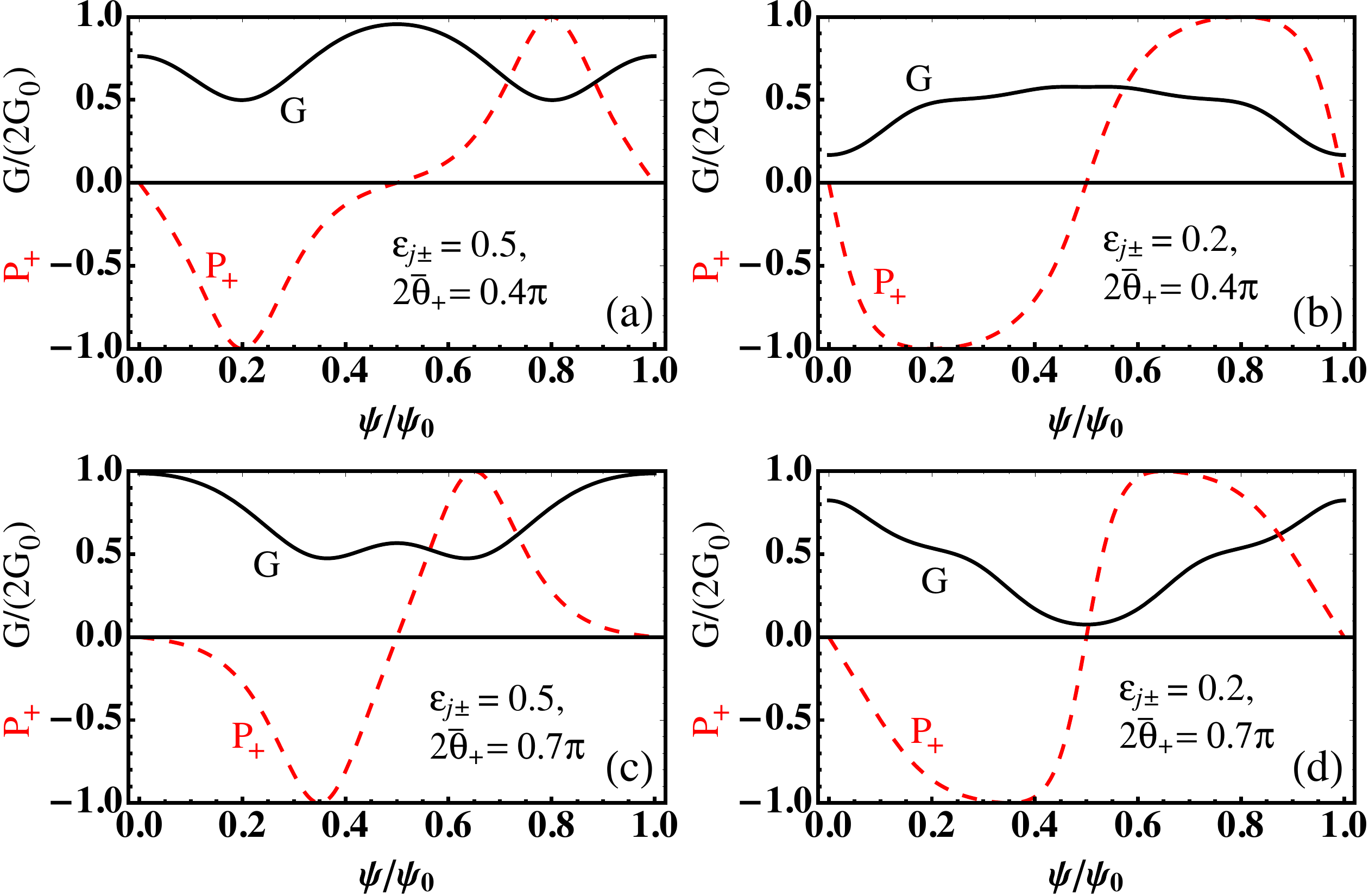}
\caption{\label{fig:bsCond}%
Dependence of the conductance $G/(2G_0)$ (black solid curves) and
$\tau$ polarization $P_+$ (red dashed curves) on magnetic flux $\psi$
(measured in units of the magnetic flux quantum $\psi_0$) for a
symmetric ring structure with beam-splitter-type T-junction \textit{S}
matrices as given, e.g., in Ref.~\onlinecite{gia11}. Other parameters used
in the calculation are $\sin^2\tilde\chi_\pm=1$, $s_{j\pm} = 1$ ($j=1,
2$), and $\varepsilon_{j\pm} = 0.5$ [$0.2$, $0.5$, $0.2$], $2\bar
\theta_+ = 0.4\,\pi$ [$0.4\,\pi$, $0.7\,\pi$, $0.7\,\pi$] for panel
(a) [(b), (c), (d)].}
\end{figure}

A T junction acting as a beam splitter is described by an \textit{S}
matrix of the form given in Eq.~(\ref{eq:genSmat}) where $|\kappa_{j
\tau}| = |\lambda_{j\tau}| \equiv \sqrt{(1 - \varepsilon_{j\tau})/2}$.
Assuming again $\chi=0$, $\varepsilon_{j\tau}\equiv\varepsilon$, and
$\sin^2\tilde\chi_\tau=1$, Eq.~(\ref{eq:tauTrans}) yields
\begin{eqnarray}
&& T_\tau^\mathrm{(bs)}\big (\theta_\mathrm{AA}^{(\tau)}\big) =
\nonumber \\ && \hspace{0.5cm}
\frac{2\varepsilon^2 \big(1 + \cos\theta_\mathrm{AA}^{(\tau)}
\big)}{\varepsilon^2 \left\{ \frac{1}{2}\big[ 1 + \cos
\theta_\mathrm{AA}^{(\tau)}\big] + 1 \right\}^2 + \frac{1-2
\varepsilon}{4}\big[ 1 - \cos\theta_\mathrm{AA}^{(\tau)}\big]^2} 
\,\,\, . \quad \nonumber \\
\end{eqnarray}
Figure~\ref{fig:bsCond} shows the magnetic-flux dependence of $G$
and $P_+$ for this case, using the same values for other parameters
as we did for the case of real T-junction \textit{S} matrices in
Fig.~\ref{fig:realCond}. Note that, for fully transparent T junctions
($\epsilon_{j\tau}=1/2$), the cases of real and beam-splitter-type
\textit{S} matrices yield identical results, as is visible from the
direct comparison of Figs.~\ref{fig:realCond}(a,c) with
Figs.~\ref{fig:bsCond}(a,c). In contrast, for a ring that is weakly
connected to leads, the different T-junction types are associated with
very different behavior. Unlike the situation with real \textit{S}
matrices, the magnitude of the flavor ($\tau$) polarization of the
current is close to unity over a significant range of values for the
magnetic flux $\psi$ in the configuration with beam-splitter T
junctions. Also in contrast to the real-\textit{S}-matrix case, the
range of flux values for maximum valley polarizations coincides with
sizable values of total conductance $G$. The general location of
valley-polarization maxima remains tunable overall by adjusting $\bar
\theta_+$, but their flux dependence does not exhibit a narrow
resonance-like line shape. Quantum-ring structures where T junctions
are of beam-splitter type thus lend themselves for use as very
effective Dirac-electron-flavor filters.

\section{Conclusions}
\label{sec:concl}

We have developed a general framework for describing theoretically
both the radial quantum confinement and the azimuthal motion of
2D-Dirac-like charge carriers in ring conductors. The formalism
applies to a wide range of 2D materials, including narrow-gap
semiconductor quantum wells as well as few-atom-thick crystals, and
also covers situations with band inversion. We present a generally
valid effective model for the azimuthal motion that illuminates a
number of interesting features of the Dirac-quantum-ring subband
structure and also yields quantitative expressions for corresponding
magnitudes. One such interesting feature is the  dependence of
subband dispersions on the flavor degree of freedom carried by
2D-Dirac-like charge carriers. Another is the competition between size
quantization and band inversion that shifts the topological regime for
the quantum-ring system to values of the 2D-Dirac gap $\Delta_0$ that
need to satisfy $\Delta_0/2 < - \gamma/W$ in terms of the ring width
$W$ and 2D-Dirac-electron Fermi velocity $\gamma/\hbar$. A
massless-Dirac-like dispersion can occur for ring-confined charge
carriers from the lowest pair of conduction and valence subbands in
the limit of large negative 2D-Dirac gap, i.e., when $\Delta_0 < 0$
and $|\Delta_0| \gg 2\gamma/W$. More generically, size-quantization
effects ensure that the subband dispersions are gapped, even for the
lowest subband, and even when the 2D-Dirac gap vanishes as is the
case, e.g., in graphene.

We have used the insight gained from calculated Dirac-ring subband
dispersions and eigenstates to investigate quantum-interference
effects in the two-terminal conductance. Our analysis is based on the
scattering-matrix approach and carefully incorporates effects arising
from the coupling to external leads. We obtain a fully general
expression for the transmission function [Eq.~(\ref{eq:tauTrans})] for
the case where T junctions with the leads conserve the charge
carriers' flavor degree of freedom and motion in the ring segments
between the leads is ballistic. Our consideration of this situation
is motivated by the recent realization of ballistic ring structures
in 2D-Dirac materials~\cite{kon06, dau17}. Properties of the
Dirac-ring subband structure turn out to uniquely influence the
flavor-dependent geometric (Berry, Aharonov-Anandan) phase
$\theta_\mathrm{AA}^{(\tau)}$, which is also dependent on magnetic
flux $\psi$ [as per Eq.~(\ref{eq:fluxAAphase})] but otherwise entirely
robust against nonuniversal, hard-to-determine experimentally, details
such as shifts in quantum phases associated with the T junctions and
the Fermi energy in the leads. Distinctive interference patterns
emerge in the two-terminal conductance that manifest unique properties
of quantum-ring subbands, including the transition between
massless-Dirac and Schr\"odinger-like behavior for the lowest one. In
addition, the dependence of interference effects on the charge
carriers' flavor degree of freedom enables use of quantum rings as
tunable flavor-filter devices. As one of the most relevant possible
sources of nonuniversal effects in experiments, we considered
variations in the design of T junctions between the ring and external
leads, including their reduced transparency. Such insight is
particularly useful to inform proper analysis of features associated
with the crossover between Dirac-like and Schr\"odinger-like behavior
expected for the lowest ($n=\pm 1$) ring subbands.

Results presented here could be further applied, or suitably
generalized, to study transport through Dirac-ring conductors that
are tunnel-coupled to leads and therefore do not conserve the charge
carriers' flavor degree of freedom. The effect of disorder scattering
in ring segments connecting the leads could similarly be investigated.
As the recently noted~\cite{yin17} remarkable robustness of persistent
currents in Dirac rings against disorder was attributed to special
properties of the lowest ring subband, we expect our two-terminal
transport results to be similarly robust.

The formalism employed in our work provides a new tool for
investigating more broadly the effect of quantum confinement on
particles whose dynamics is governed by a Dirac equation. It could be
usefully applied to related effective-model descriptions of charge
carriers in semi-metallic systems such as Weyl~\cite{yan17b} and
nodal-line~\cite{fan16} semimetals, opening up new possibilities to
explore quantum-transport effects in these, and similar, topological
materials. Another interesting avenue for future studies expanding on
our approach is Dirac-electron physics in hybrid structures.
Situations of this type have been considered before in tight-binding
transport calculations for graphene rings with superconducting
leads~\cite{sch12a} or subject to electrostatic
potentials~\cite{sch10}.

\begin{acknowledgments}

L.G.\ is the grateful recipient of a Victoria University of Wellington
Masters by thesis Scholarship. U.Z.\ thanks P.M.R.~Brydon for useful
discussions. R.W.\ appreciates the hospitality of Victoria University
of Wellington during research visits while this work was performed, as
well as support by the NSF under grant No.\ DMR-1310199. Work at
Argonne was supported by the U.S. Department of Energy, Office of
Science, Basic Energy Sciences, Materials Science and Engineering
Division under Contract No.\ DE-AC02-06CH11357. 

\end{acknowledgments}

\appendix

\section{Determination of hard-wall-confined quantum-ring bound
states with \texorpdfstring{\bm{$l=0$}}{}}\label{sec:appA}

To simplify the notation, we introduce the dimensionless quantities
$\tilde k = k/k_\Delta$, $\tilde E \equiv E/(\gamma k_\Delta)$, and
${\tilde\Delta}_0 = \Delta_0/(\gamma k_\Delta)$. The energy
eigenvalues and corresponding (non-normalized) eigenstates of the
Hamiltonian $\mathcal{H}_\mathrm{1D}^{(\tau)}$ can then be expressed
as
\begin{subequations}
\begin{eqnarray}
{\tilde E}_{k \pm} &=& \xi\, {\tilde k}^2 \pm \sqrt{{\tilde
k}^4 + \left( {\tilde\Delta}_0 + 1 \right) {\tilde k}^2 +
\frac{{\tilde\Delta}_0^2}{4}} \quad , \quad \\[0.2cm]
\Phi_{k \pm}^{(\tau)}(r) &=&  \left( \begin{array}{c} 1 \\ \pm \tau
\, \mathrm{sgn}(k) \, \gamma_{k \pm} \end{array}\right) \ee^{i k r}
\quad ,
\end{eqnarray}
with the abbreviation
\begin{equation}
\gamma_{k \pm} = \sqrt{\frac{{\tilde E}_{k \pm} - ({\tilde\Delta}_0
/ 2) - (1+ \xi) {\tilde k}^2}{{\tilde E}_{k \pm} + ({\tilde
\Delta}_0 / 2) + (1 - \xi) {\tilde k}^2}} \quad .
\end{equation}
\end{subequations}

Solutions of the confinement problem (\ref{eq:l=0SE}) with a
hard-wall potential (\ref{eq:hardwall}) are found by forming a
general superposition of the possible eigenstates of
$\mathcal{H}_\mathrm{1D}^{(\tau)}$ with given energy $\tilde E$ and
imposing hard-wall boundary conditions. Focusing initially on
$|\tilde E| \ge |{\tilde\Delta}_0|/2$, four wave numbers are obtained
as roots of the equation ${\tilde E}_{k\pm} = \tilde E$. Two of them
are real and given by $\pm\tilde k$, the other two are
imaginary and given by $\pm i \tilde q$. We find the explicit
expressions~\footnote{The existence of these solutions is guaranteed
for any combination of materials parameters as long as $|\xi|<1$ and
$k_\Delta > 0$. These conditions are indeed satisfied for the
materials of interest here.}
\begin{widetext}
\begin{subequations}
\begin{eqnarray}\label{eq:kConf}
\tilde k &=& \left[ \frac{\sqrt{({1+ \tilde\Delta}_0 + 2\xi
\, \tilde E)^2 + (1 - \xi^2)(4 {\tilde E}^2 - {\tilde
\Delta}_0^2)} - (1 + {\tilde\Delta}_0 + 2\xi\, \tilde E)}{2
(1 - \xi^2)}\right]^{\frac{1}{2}} \quad , \\[0.2cm]
\tilde q &=& \left[ \frac{\sqrt{(1 + {\tilde\Delta}_0 + 2
\xi\, \tilde E)^2 + (1 - \xi^2)(4 {\tilde E}^2 -
{\tilde\Delta}_0^2)} + 1 + {\tilde\Delta}_0 + 2\xi\,
\tilde E}{2 (1 - \xi^2)}\right]^{\frac{1}{2}} \quad .
\end{eqnarray}
\end{subequations}
The full \textit{Ansatz\/} for the bound-state wave function is
\begin{equation}\label{eq:BHZstate}
\Phi_0^{(\tau, n)}(r) = c_{1 k}^{(\tau, n)} \left( \begin{array}{c}
1 \\ \tau \, \gamma_k \end{array}\right) \ee^{i k r} +\, c_{2
k}^{(\tau,n)} \left( \begin{array}{c} 1 \\ - \tau\, \gamma_k
\end{array}\right) \ee^{- i k r} + \, c_{1 q}^{(\tau, n)} \left(
\begin{array}{c} {\bar \gamma}_q \\ - i \tau \end{array} \right)
\ee^{- q r} +\, c_{2 q}^{(\tau, n)} \left( \begin{array}{c}
{\bar\gamma}_q \\ i \tau \end{array}\right) \ee^{q r} \quad ,
\end{equation}
with the parameters
\begin{subequations}
\begin{eqnarray}
\gamma_k &=& \mathrm{sgn}(\tilde E)\, \sqrt{\frac{1+\xi}{1-\xi}}\,
\left[\frac{2 \tilde E + \xi\, {\tilde\Delta}_0 - \sqrt{(1 + {\tilde
\Delta}_0 + 2\xi\,\tilde E)^2 + (1 - \xi^2)(4 \tilde E^2 - {\tilde
\Delta}_0^2)} + 1}{2\tilde E + \xi \, {\tilde\Delta}_0 + \sqrt{(1 +
{\tilde\Delta}_0 + 2 \xi\,\tilde E)^2 + (1 - \xi^2)(4 \tilde E^2 -
{\tilde\Delta}_0^2)} - 1}\right]^{\frac{1}{2}} \quad , \\[0.2cm]
{\bar\gamma}_q &=& \sqrt{\frac{1-\xi}{1+\xi}}\, \left[ \frac{\sqrt{(
1 + {\tilde\Delta}_0 + 2\xi\,\tilde E)^2 + (1 - \xi^2)(4 \tilde E^2
- {\tilde\Delta}_0^2)} + 1 - 2 \tilde E - \xi\, {\tilde
\Delta}_0}{\sqrt{(1 + {\tilde\Delta}_0 + 2\xi \,\tilde E)^2
+ (1 - \xi^2)(4 \tilde E^2 - {\tilde\Delta}_0^2)} + 1 + 2
\tilde E + \xi \, {\tilde\Delta}_0}\right]^{\frac{1}{2}} \quad .
\end{eqnarray}
\end{subequations}
\end{widetext}
The secular equation obtained from imposing hard-wall boundary
conditions $\Phi_0^{(\tau, n)}(R\pm W/2) = 0$ at the inner and outer
ring radii is similar to those found in related bound-state
problems~\cite{whi81, mic11}. It can be written in the concise form
\begin{equation}\label{eq:BHZ_sec_ab}
\gamma_k \, {\bar\gamma}_q = \left\{ \begin{array}{rl} \tanh (q W/2 )
\cot (k W/2 ) & \hspace{0.2cm} \mbox{case a} \, , \\[0.3cm] - \coth
( q W/2 ) \tan ( k W/2 ) & \hspace{0.2cm}\mbox{case b}\, ,
\end{array} \right.
\end{equation}
where case $\nu=\mathrm{a}$ ($\nu=\mathrm{b}$) yields solutions with
even (odd) parity associated with eigenvalues $E_0^{(\tau,n_\nu)}$.
The corresponding eigenstates can be written as
\begin{equation}
\Phi_0^{(\tau, n_\nu)}(r) = \Phi_\mathrm{D}^{(\tau, n_\nu)}(r) +
\Phi_\mathrm{B}^{(\tau, n_\nu)}(r) \quad ,
\end{equation}
where the contribution labeled D has the form of a standing-wave
state for a Dirac particle~\cite{alb96}, and the part labeled B is
an evanescent correction that incorporates the remote-band
contributions~\cite{whi81}. More explicitly, we find
\begin{subequations}\label{eq:DiracWing}
\begin{eqnarray}\label{eq:BHD_Dirac_1}
\Phi_\mathrm{D}^{(\tau, n_\mathrm{a})}(r) &=&
{\mathcal N}_{n_\mathrm{a}} \left( \begin{array}{c} \cos\left[
k_{n_\mathrm{a}} (r - R)\right] \\[0.2cm] \tau\,
\gamma_{k_{n_\mathrm{a}}} \, i \sin\left[ k_{n_\mathrm{a}} (r - R)
\right] \end{array} \right)\, ,\quad \\[0.2cm] \label{eq:BHD_wing_1}
\Phi_\mathrm{B}^{(\tau, n_\mathrm{a})}(r) &=& -
{\mathcal N}_{n_\mathrm{a}}\, \gamma_{k_{n_\mathrm{a}}}\,
\frac{\sin (k_{n_\mathrm{a}} W/2 )}{\sinh (q_{n_\mathrm{a}} W/2)}
\nonumber \\[0.1cm] && \hspace{0.5cm} \times\left( \begin{array}{c}
{\bar\gamma}_{q_{n_\mathrm{a}}} \cosh\left[ q_{n_\mathrm{a}}(r - R)
\right] \\[0.2cm] \tau\, i \sinh\left[ q_{n_\mathrm{a}}(r - R)
\right] \end{array} \right) \, , \quad \\[0.2cm]
\label{eq:BHD_Dirac_2} \Phi_\mathrm{D}^{(\tau, n_\mathrm{b})}(r) &=&
{\mathcal N}_{n_\mathrm{b}} \left( \begin{array}{c} \sin\left[
k_{n_\mathrm{b}} (r - R)\right] \\[0.2cm] - \tau\,
\gamma_{k_{n_\mathrm{b}}} \, i \cos\left[ k_{n_\mathrm{b}} (r - R)
\right] \end{array} \right) , \quad \\[0.2cm] \label{eq:BHD_wing_2}
\Phi_\mathrm{B}^{(\tau, n_\mathrm{b})}(r) &=& {\mathcal
N}_{n_\mathrm{b}}\, \gamma_{k_{n_\mathrm{b}}}\, \frac{\cos (
k_{n_\mathrm{b}} W/2 )}{\cosh ( q_{n_\mathrm{b}} W/2 )} \nonumber
\\[0.1cm] && \hspace{0.5cm} \times\left( \begin{array}{c} {\bar
\gamma}_{q_{n_\mathrm{b}}} \cosh \left[ q_{n_\mathrm{b}} (r - R)
\right] \\[0.2cm] \tau\, i \sinh\left[ q_{n_\mathrm{b}} (r - R)
\right] \end{array}\right) \, , \end{eqnarray}
\end{subequations}
where the ${\mathcal N}_{n_\nu}$ denote normalization factors.

Previous results~\cite{whi81, ber87, mck87, cou88, alb96} for Dirac
particles with hard-wall mass confinement are reproduced in the limit
$\Delta(k)\to\Delta_0$ and $\epsilon(k)\to 0$, which is achieved by
taking $k_\Delta\to\infty$ and $\xi\to 0$ in all relevant
expressions~\footnote{It turns out that both limits $k_\Delta\to
\infty$ and $\xi\to 0$ are required to ensure $\bar\gamma_q \to 1$.
This is an intricacy arising from the well-known~\cite{whi81, sch85}
(and otherwise for our purposes benign) properties of the evanescent
`wing' states with imaginary wave vectors $\pm iq$ in the
\textit{Ansatz\/} (\ref{eq:BHZstate}). See also Ref.~\onlinecite{kli17} for
a related discussion}. In the process, we have $q\to\infty$, $k\to
\sqrt{4 E^2 - \Delta_0^2}/(2\gamma)$, $\gamma_k\to\mathrm{sgn}(E)
\sqrt{(2 E - \Delta_0)/(2 E + \Delta_0)}$, and $\bar\gamma_q\to 1$. As
a result, the secular equation (\ref{eq:BHZ_sec_ab}) simplifies
to~\cite{whi81}
\begin{equation}
\gamma_k  = \left\{ \begin{array}{rl} \cot ( k W/2 ) &
\hspace{0.2cm}\mbox{case a}\, , \\[0.2cm] - \tan ( k W/2 )
& \hspace{0.2cm}\mbox{case b}\, , \end{array} \right.
\end{equation}
eigenstates are purely of the Dirac-standing-wave form $\Phi_0^{(
\tau, n_\nu)}(r) \to \Phi_\mathrm{D}^{(\tau, n_\nu)}(r)$, and the
normalization factors are given by
\begin{equation}
{\mathcal N}_{n_\nu} = \frac{1}{\sqrt{W}} \left[\frac{E_0^{(\tau,
n_\nu)} \left( 2 E_0^{(\tau,n_\nu)} + \Delta_0 \right)}{2 \left(
E_0^{(\tau,n_\nu)}\right)^2 + E_W\, \Delta_0} \right]^{\frac{1}{2}} ,
\end{equation}
where $E_W \equiv \gamma/W$ is the energy scale associated with
size quantization for the confined Dirac particles.

\begin{figure}[t]
\includegraphics[width=\columnwidth]{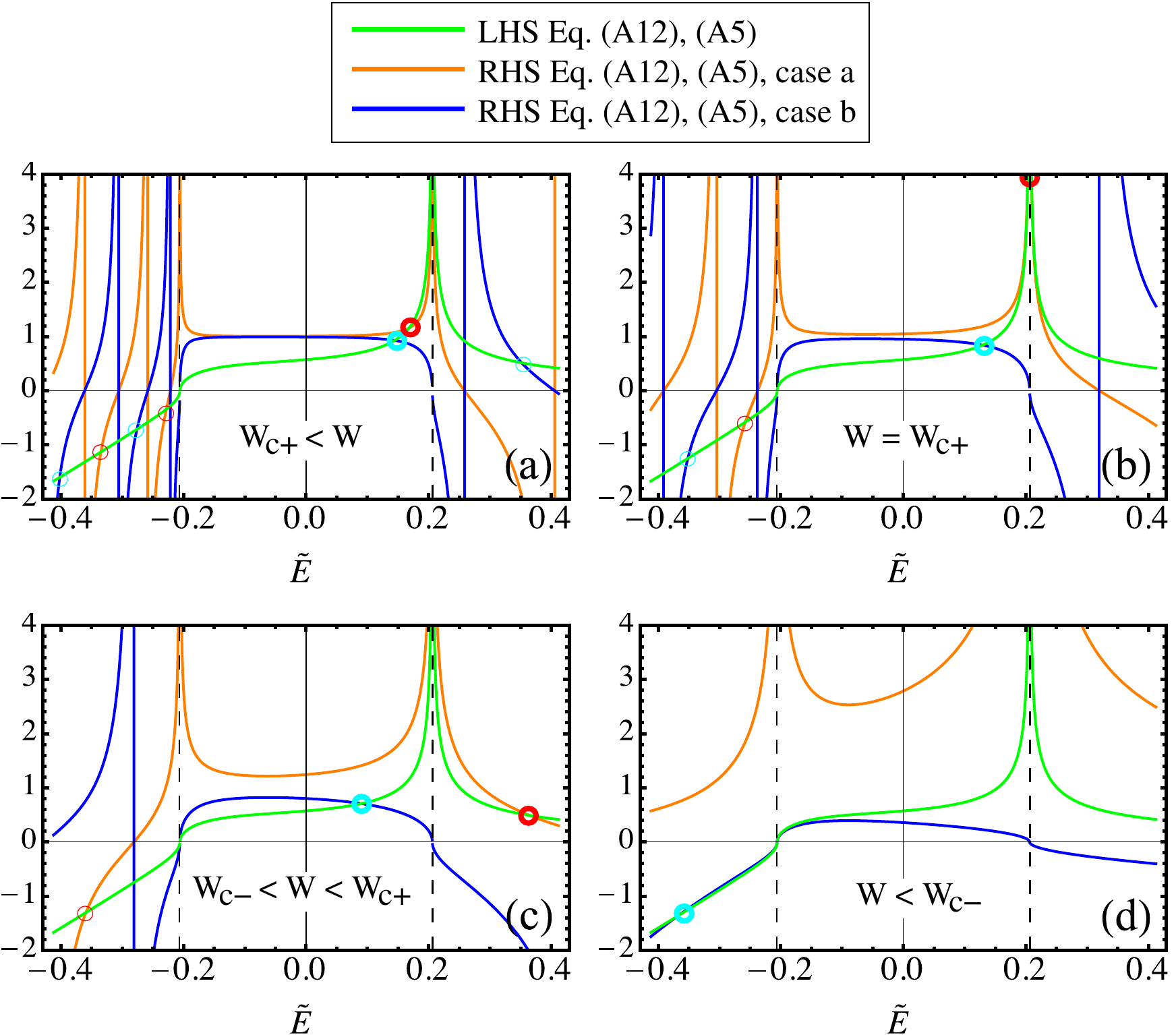}
\caption{\label{fig:BHZspec}%
Ring-confined Dirac-particle states with $l=0$ related to
quantum-spin-Hall edge states. (a) Two bound states with energy
inside the gap exist for sufficiently large widths $W_{\mathrm{c}+} <
W$. (b) When $W = W_{\mathrm{c}+}$, one of these states is pushed
through the top of the gap. (c) For $W_{\mathrm{c}-} < W <
W_{\mathrm{c}+}$, only one bound-state energy is still within the
gap. (d) No bound states exist within the gap for $W <
W_{\mathrm{c}-}$. In all panels (a) to (d), the LHS expression of the
secular equation (\ref{eq:evanesSE}) [(\ref{eq:BHZ_sec_ab})] is
represented by the green curve in the region $|\tilde E| < |{\tilde
\Delta}_0|/2$ [$|\tilde E| > |{\tilde\Delta}_0|/2$], while the RHS
for case a (case b) is plotted as the orange (blue) curve. Parameters
used are ${\tilde\Delta}_0 = -0.412$, $\xi = 0.746$ (corresponding to
a 7-nm HgTe quantum well~\cite{rot10}), and $k_\Delta W = 20$, $13.4$,
$8$, $2.3$.}
\end{figure}

So far, we have considered bound states of ring-confined Dirac
particles that are typical standing waves, i.e., are extended in
radial direction across the ring. However, it is well-known that the
presence of a band inversion (signified within our model by $\Delta_0
< 0$) gives rise to topologically protected states localized at the
system boundaries~\cite{vol85, kan05, zho08, win17}, which should also
appear in our situation of interest~\cite{mic11}. In fact, our
\textit{Ansatz\/} (\ref{eq:BHZstate}) applies to energies $|\tilde
E| < |{\tilde\Delta}_0|/2$ if the replacements 
\begin{equation}\label{eq:evanescRel}
k \equiv i {\bar k} \,\,\,\, \mbox{ and }\,\,\,\, \gamma_k \equiv
i\, {\bar\gamma}_{\bar k} \quad ,
\end{equation}
are made, with the real quantities~\footnote{To exclude situations
where $\bar k$ and $\bar\gamma_{\bar k}$ are complex, we assume
the condition~\cite{zho08, mic11} $\tilde\Delta_0 > -1/[2(1-\xi^2)]$
to be satisfied.}
\begin{widetext}
\begin{subequations}
\begin{eqnarray}\label{eq:kEvConf}
\tilde{\bar k} &=& \left[ \frac{1 + {\tilde\Delta}_0 + 2\xi
\,\tilde E - \sqrt{(1 + {\tilde\Delta}_0 + 2\xi \, \tilde
E)^2 + (1 - \xi^2)(4 \tilde E^2 - {\tilde\Delta}_0^2)}}{2
(1 - \xi^2)}\right]^{\frac{1}{2}} \quad , \\[0.2cm]
{\bar\gamma}_{\bar k} &=& \mathrm{sgn}({\tilde\Delta}_0) \,
\sqrt{\frac{1+\xi}{1-\xi}}\,\,\left[ \frac{\sqrt{(
1 + {\tilde\Delta}_0 + 2\xi\,\tilde E)^2 + (1 - 
\xi^2)(4 \tilde E^2 - {\tilde\Delta}_0^2)} - 1 - 2 \tilde E
- \xi\, {\tilde\Delta}_0}{\sqrt{( 1 + {\tilde\Delta}_0 + 2
\xi\,\tilde E)^2 + (1 - \xi^2)(4 \tilde E^2 -
{\tilde\Delta}_0^2)} - 1 + 2 \tilde E + \xi \, {\tilde
\Delta}_0}\right]^{\frac{1}{2}} \quad .
\end{eqnarray}
\end{subequations}
\end{widetext}
The secular equation for $|\tilde E| < |{\tilde\Delta}_0|/2$ then
reads~\cite{zho08}
\begin{equation}\label{eq:evanesSE}
- {\bar\gamma}_{\bar k} \, {\bar\gamma}_q = \left\{ \begin{array}{rl}
\tanh ( q W/2 ) \coth ( {\bar k} W/2 ) & \hspace{0.2cm} \mbox{case a}
\, , \\[0.2cm] \coth ( q W/2 )\tanh ( {\bar k} W/2 ) & \hspace{0.2cm}
\mbox{case b}\, . \end{array} \right.
\end{equation}
Corresponding eigenstates for the in-gap bound states are obtained by
using the expressions from Eq.~(\ref{eq:evanescRel}) in the wave
functions shown in Eqs.~(\ref{eq:DiracWing}a-d). Because of the
positivity of its RHS, solutions of Eq.~(\ref{eq:evanesSE}) exist
only for $\Delta_0 < 0$, and there can be at most one solution for
each case a and b. Figure~\ref{fig:BHZspec} illustrates the regimes
for which two, one, or no bound states have energies within the gap,
assuming materials parameters for a 7-nm HgTe quantum well. The
boundaries between these regimes in parameter space can be
associated with critical ring widths $W_{\mathrm{c}\pm}$. We find
analytical expressions in the typical situation where $q W\gg 1$;
\begin{subequations}
\begin{eqnarray}
W_{\mathrm{c}\pm} &=& \lim_{E\to \mp \frac{\Delta_0}{2}} \left\{
\frac{2}{\bar k} \, \mathrm{arcoth}\left(\left[{\bar\gamma}_{\bar k}
\,{\bar\gamma}_q \right]^{\pm 1} \right)\right\} \quad , \\[0.2cm]
&\equiv& \frac{2}{|{\tilde\Delta}_0|} \left[\frac{1 \mp \xi}{(1 \pm
\xi)[1 + ( 1 \pm \xi ) {\tilde\Delta}_0]} \right]^\frac{1}{2} 
k_\Delta^{-1} . \quad
\end{eqnarray}
\end{subequations}
The critical ring widths arise due to the fact that the
size-quantization energy reduces the magnitude of the negative
(topological) gap parameter, thereby driving a transition from the
topological into the normal regime for the ring band structure that
is analogous to similar transitions in higher-dimensional
systems~\cite{ber06, liu10, kot17}. Note also that, in the limit $W\to
\infty$, the secular equations (\ref{eq:evanesSE}) read $\bar
\gamma_{\bar k} \, {\bar\gamma}_q = -1$, which has the
solution~\cite{zho08, son10} $\tilde E = -\xi\, {\tilde\Delta}_0/2$.

\section{Derivation of the effective Dirac-ring Hamiltonian in the
\texorpdfstring{\bm{$l=0$}}{}-bound-state basis}\label{sec:appB}

The azimuthal motion of ring-confined Dirac particles is described
by Eq.~(\ref{eq:xLham}). Here we analyze the structure of its $2
\times 2$ sub-block matrices (\ref{eq:2x2Lham}).

The diagonal blocks having $n=n'$ are Hermitian matrices and can
therefore be written as a superposition of Pauli matrices $\eta_j$
that are acting in the $2\times 2$ subspace spanned by $l=0$
eigenstates $\ket{\Phi_0^{(\tau, \pm n)}}$,
\begin{equation}
\left( \ham_l^{(\tau)} \right)_{n, n} = \sum_{j = 0}^3
\Gamma_j^{(\tau,n)}(l) \,\, \eta_j \quad .
\end{equation}
The most general expression for the coefficients
$\Gamma_{l j}^{(\tau,n)}$ are
\begin{subequations}
\begin{eqnarray}
\Gamma_{0}^{(\tau,n)}(l) &=& \frac{1}{2} \Big( E_0^{(\tau,n)} +
E_0^{(\tau, -n)} \nonumber \\ && \hspace{0.3cm} +\, \big\langle
\mathcal{V}^{(\tau)}_l(r) \big\rangle^{(\tau)}_{n,n} + \big\langle
\mathcal{V}^{(\tau)}_l(r)\big\rangle^{(\tau)}_{-n,-n}\Big) , \quad
\\[0.2cm] \Gamma_1^{(\tau,n)}(l) &=& \Re\mathrm{e}\left\{ \big\langle
\mathcal{V}^{(\tau)}_l(r) \big\rangle^{(\tau)}_{n,-n} \right\} \quad ,
\\[0.2cm] \Gamma_2^{(\tau,n)}(l) &=& -\Im\mathrm{m}\left\{ \big
\langle\mathcal{V}^{(\tau)}_l(r) \big\rangle^{(\tau)}_{n,-n} \right\} \quad ,
\\[0.2cm] \Gamma_3^{(\tau,n)}(l) &=& \frac{1}{2} \Big( E_0^{(\tau,
n)} - E_0^{(\tau,-n)} \nonumber \\ && \hspace{0.3cm} +\, \big\langle
\mathcal{V}^{(\tau)}_l(r)\big\rangle^{(\tau)}_{n,n} - \big\langle
\mathcal{V}^{(\tau)}_l(r) \big\rangle^{(\tau)}_{-n,-n}\Big) . \quad
\end{eqnarray}
\end{subequations}
These expressions simplify considerably in the
electron-hole-symmetric situation $\xi = 0$;
\begin{subequations}\label{eq:effRingHam}
\begin{eqnarray}
\Gamma_{0,\xi=0}^{(\tau,n)}(l) &=& \gamma\, l \left[ \big\langle
\sigma_2/r \big\rangle^{(\tau)}_{n,n} - \tau\, \big\langle \sigma_0 /
(k_\Delta r^2)\big\rangle^{(\tau)}_{n,n} \right] , \quad \\[0.2cm]
\Gamma_{1,\xi=0}^{(\tau,n)}(l) &=& \gamma\, l \left[ \big\langle
\sigma_0/r \big\rangle^{(\tau)}_{n,n} - \tau\, \big\langle \sigma_2 /
(k_\Delta r^2)\big\rangle^{(\tau)}_{n,n} \right] , \quad \\[0.2cm]
\Gamma_{2,\xi=0}^{(\tau,n)}(l) &=& \gamma\, l^2 \, \big\langle
\sigma_1/(k_\Delta r^2) \big\rangle^{(\tau)}_{n,n} \quad , \\[0.2cm]
\Gamma_{3,\xi=0}^{(\tau,n)}(l) &=& E_{0,\xi=0}^{(\tau,n)} +\gamma\,
l^2 \, \big\langle \sigma_3 /(k_\Delta r^2)\big\rangle^{(\tau)}_{n,n}
\quad .
\end{eqnarray}
\end{subequations}
Treating the electron-hole-asymmetry contribution $\propto\xi$
in $\mathcal{V}^{(\tau)}_l(r)$ perturbatively, one can approximate
$\Gamma_j^{(\tau,n)}(l)\approx\Gamma_{j,\xi=0}^{(\tau,n)}(l) +
\delta\Gamma_j^{(\tau,n)}(l)$, with corrections given to first
order in small $\xi$ by
\begin{subequations}\label{eq:effRingHamCorr}
\begin{eqnarray}
\delta\Gamma_0^{(\tau,n)}(l) &=& \xi\, \big( \partial E_0^{(\tau,n)}
/\partial \xi \big)_{\xi = 0} + \xi\, \gamma\, l^2 \, \big\langle
\sigma_0/(k_\Delta r^2) \big\rangle^{(\tau)}_{n,n} \, , \nonumber
\\ \\[0.2cm] \delta\Gamma_1^{(\tau,n)}(l) &=& \xi\, \gamma\, l^2 \,
\big\langle\sigma_2/(k_\Delta r^2) \big\rangle^{(\tau)}_{n,n} \quad ,
\\[0.2cm] \delta\Gamma_2^{(\tau,n)}(l) &=& \xi\, \gamma\, (-\tau) \, l
\, \big\langle\sigma_1/(k_\Delta r^2) \big\rangle^{(\tau)}_{n,n} \quad ,
\\[0.2cm] \delta\Gamma_3^{(\tau,n)}(l) &=&  \xi\, \gamma\, (-\tau) \, l
\, \big\langle\sigma_3/(k_\Delta r^2) \big\rangle^{(\tau)}_{n,n} \quad .
\end{eqnarray}
\end{subequations}

Further analysis is facilitated by substituting the eigenstates for a
hard-wall mass confinement from Eqs.~(\ref{eq:DiracWing}) as the
states between which matrix elements in Eqs.~(\ref{eq:effRingHam}a-d)
and (\ref{eq:effRingHamCorr}a-d) are calculated. It is then useful to
define the quantities
\begin{equation}
\Xi_{j m}^{(\tau,n)}(W/R) = \big\langle \sigma_j \, (W/r)^m \big
\rangle^{(\tau)}_{n,n} \quad ,
\end{equation}
as these are functions of the ring aspect ratio $W/R$. The natural
energy scale of the $\Gamma_j^{(\tau,n)}(l)$ is then $E_W$, and
terms quadratic in $l$ are suppressed by a factor $1/(k_\Delta W)
\ll 1$ typically. From the particular form and $r$ dependence of the
spinors in Eqs.~(\ref{eq:DiracWing}), it can be deduced that the
leading-order behavior in $W/R$ is $\Xi_{j m}^{(\tau,n)}(W/R)\sim
(W/R)^m$ for $j=0$ and $j=3$ (i.e., the matrix elements involving
diagonal Pauli matrices) whereas $\Xi_{j m}^{(\tau, n)}(W/R) \sim
(W/R)^{m+1}$ for $j=1$ and $j=2$. Also, the functions $\Xi_{j
m}^{(\tau,n)}$ with $j=0$ and $3$ ($j=1$ and $2$) are independent
of (proportional to) $\tau$. Finally, $\Xi_{1 m}^{(\tau,n)}(W/R)\equiv
0$ identically because the upper (lower) entry in the eigenspinors
given by Eqs.~(\ref{eq:DiracWing}) is always real (imaginary).
Based on these insights, we parameterize the effective Hamiltonian
for azimuthal motion within subbands with labels $\pm n$ in the form
given in Eqs.~(\ref{eq:aziHam}), (\ref{eq:aziHam0}) and
(\ref{eq:aziHam1}).

Taking the limit $W/2 \to R$ in our model yields results that are
directly applicable to mass-confined 2D-Dirac electrons in circular
quantum dots~\cite{gut15}. However, as $W/R$ is not small in that
situation, no hierarchy of magnitudes between matrix elements $\big(
\ham_l^{(\tau)} \big)_{n, n'}$ in Eq.~(\ref{eq:xLham}) can be
established. Hence, unlike in the case of quantum rings, we cannot
obtain a simple effective Hamiltonian that accurately describes the
azimuthal motion of 2D-Dirac electrons in a quantum dot.

%

\end{document}